\documentclass[aps,prc,showpacs,twocolumn,superscriptaddress]{revtex4-1}
\usepackage{amsmath,amssymb,graphicx,bm,braket}
\usepackage[caption=false]{subfig}
\renewcommand{\vec}[1]{\mathbf{#1}}
  
\newcommand{\clebschG}[6]{\mathcal{C}_{#1 #4 #2 #5}^{#3 #6}}  
	
\begin{document}

\title{Microscopic calculations and energy expansions
for neutron-rich matter}

\author{C.\ Drischler}
\email[E-mail:~]{cdrischler@theorie.ikp.physik.tu-darmstadt.de}
\affiliation{Institut f\"ur Kernphysik, Technische Universit\"at Darmstadt, 
64289 Darmstadt, Germany}
\affiliation{ExtreMe Matter Institute EMMI, 
GSI Helmholtzzentrum f\"ur Schwerionenforschung GmbH, 64291 Darmstadt, Germany}
\author{V.\ Som\`a}
\email[E-mail:~]{vittorio.soma@physik.tu-darmstadt.de}
\affiliation{Institut f\"ur Kernphysik, Technische Universit\"at Darmstadt, 
64289 Darmstadt, Germany}
\affiliation{ExtreMe Matter Institute EMMI, 
GSI Helmholtzzentrum f\"ur Schwerionenforschung GmbH, 64291 Darmstadt, Germany}
\author{A.\ Schwenk}
\email[E-mail:~]{schwenk@physik.tu-darmstadt.de}
\affiliation{ExtreMe Matter Institute EMMI, 
GSI Helmholtzzentrum f\"ur Schwerionenforschung GmbH, 64291 Darmstadt, Germany}
\affiliation{Institut f\"ur Kernphysik, Technische Universit\"at Darmstadt, 
64289 Darmstadt, Germany}

\begin{abstract}
We investigate asymmetric nuclear matter with two- and three-nucleon
interactions based on chiral effective field theory, where three-body
forces are fit only to light nuclei. Focusing on neutron-rich matter,
we calculate the energy for different proton fractions and include
estimates of the theoretical uncertainty. We use our ab-initio results
to test the quadratic expansion around symmetric matter with the
symmetry energy term, and confirm its validity for highly asymmetric
systems. Our calculations are in remarkable agreement with an empirical
parametrization for the energy density. These findings are very useful for
astrophysical applications and for developing new equations of state.
\end{abstract}

\pacs{21.65.Cd, 21.30.-x, 26.60.Kp}

\maketitle

\section{Introduction}

Microscopic calculations of asymmetric nuclear matter are of great
importance because of applications for nuclei and nuclear
astrophysics, as well as from a general many-body theory
perspective. Nuclei along isotopic chains span a considerable range of
neutron-to-proton asymmetries, which influences many of their
properties. In astrophysical environments, the equation of state of
neutron-rich matter is key for core-collapse supernovae, neutron stars,
and mergers of compact objects. Moreover, calculations of asymmetric
matter can be used to guide nuclear energy-density functionals, in
particular for the evolution to neutron-rich systems.

While neutron matter and symmetric matter have been investigated
extensively, there are few microscopic studies of asymmetric
matter, because the phase space with different neutron and proton
Fermi seas is more involved. The first microscopic calculation with
simple interactions dates back to Brueckner, Coon, and
Dabrowski~\cite{Brueckner:1968zza}. This was followed by variational
calculations with phenomenological two- (NN) and three-nucleon (3N)
potentials~\cite{Lagaris:1981as}, Brueckner-Hartree-Fock
calculations~\cite{Bombaci:1991zz,Zuo:1999bd,Zuo:2002sg,Vidana:2009is},
Auxiliary-Field Diffusion Monte Carlo with a simplified
potential~\cite{Fantoni:2008jd}, and, at finite temperature,
self-consistent Green's function methods~\cite{Frick:2004th}.
Phenomenologically, one can also obtain information about the
properties of asymmetric matter by using a quadratic expansion to
interpolate between symmetric and neutron matter.

With the development of chiral effective field theory (EFT) to nuclear
forces~\cite{Epelbaum:2008ga} and the renormalization
group (RG)~\cite{Bogner:2009bt}, which improves the many-body convergence,
it is timely to revisit the study of asymmetric nuclear matter. Chiral
EFT provides a systematic expansion for NN, 3N, and higher-body
interactions with theoretical uncertainties. This is especially
important for calculations of neutron-rich matter. Nuclear forces
based on chiral EFT have been successfully used to study light to
medium-mass nuclei, nuclear reactions, and nuclear
matter~\cite{Hammer:2012id}. In particular, neutron matter has been
found to be perturbative for low-momentum interactions based on chiral
EFT potentials~\cite{Hebeler:2009iv} (see also
Ref.~\cite{Bogner:2005sn} for symmetric matter), and the perturbative
convergence was recently validated with first Quantum Monte Carlo
calculations for chiral EFT interactions~\cite{Gezerlis:2013ipa}. For
symmetric matter, the same low-momentum interactions predict realistic
saturation properties within theoretical uncertainties using 3N forces
fit only to light nuclei~\cite{Hebeler:2010xb}. The properties of
nucleonic matter were also studied using in-medium chiral perturbation
theory approaches~\cite{Kaiser:2001jx,Lacour:2009ej,Fiorilla:2011sr,
Holt:2013fwa}, lattice chiral EFT~\cite{Epelbaum:2008vj},
and self-consistent Green's functions~\cite{Carbone2014nm}.
Finally, neutron matter was
calculated completely to N$^3$LO including NN, 3N, and 4N
interactions~\cite{Tews:2012fj,Kruger:2013kua}.

In this paper, we present the first calculations of asymmetric nuclear
matter with NN and 3N interactions based on chiral EFT, which are fit
only to few-body data. We focus on neutron-rich conditions and present
results for the energy of asymmetric matter with proton fractions $x
\leqslant 0.15$. In Sect.~\ref{sec:scheme}, we discuss the NN and 3N
interactions used, outline the calculational strategy, and give the
different interaction contributions in asymmetric
matter. Section~\ref{sec:energy} presents our ab-initio results for
the energy of asymmetric matter, which we use to test the quadratic
expansion and the symmetry energy in Sect.~\ref{sec:parabolic}.  In
Sect.~\ref{sec:expansion}, we study an empirical parametrization of
the energy, which was used in Ref.~\cite{Hebeler:2013nz} to extend
ab-initio calculations of neutron matter to asymmetric matter for
astrophysical applications. Finally, we conclude in
Sect.~\ref{sec:conclusion}.

\section{Formalism}
\label{sec:scheme}

\subsection{Nuclear Hamiltonian} 
\label{sec:H}

We consider nuclear matter as an infinite, homogeneous system of
neutrons and protons governed by a many-nucleon Hamiltonian
\begin{equation}
H(\Lambda) = T + V_\text{NN}(\Lambda) + V_\text{3N}(\Lambda) + \ldots \,,
\label{eq:Hamiltonian}
\end{equation}
which depends on a resolution scale $\Lambda$. In this work, we
include NN and 3N interactions based on chiral
EFT~\cite{Epelbaum:2008ga,Machleidt:2011zz}. To improve the many-body
convergence~\cite{Bogner:2009bt}, we evolve the N$^3$LO $500 \, {\rm
MeV}$ NN potential of Ref.~\cite{Entem:2003ft} to low-momentum
interactions $V_{{\rm low}\,k}$ with a resolution scale $\Lambda = 1.8
- 2.8 \, {\rm fm}^{-1}$ and a smooth $n_\text{exp}=4$
regulator~\cite{Bogner:2006vp}. This follows the calculations of
neutron and symmetric nuclear matter of
Refs.~\cite{Hebeler:2009iv,Hebeler:2010xb}.

At the 3N level, we include the leading N$^2$LO 3N
forces~\cite{vanKolck:1994yi,Epelbaum:2002vt}, which consist of a
long-range two-pion-exchange part $V_c$ (with $c_i$ couplings), an
intermediate-range one-pion-exchange part $V_D$, and a short-range 3N
contact interaction $V_E$:
\begin{center}
\includegraphics[width=0.375\textwidth]{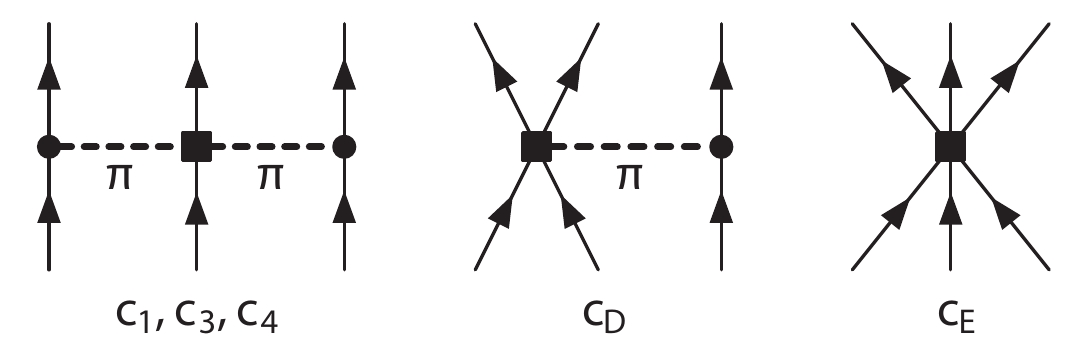}
\end{center}
Their structures are given explicitly in Appendix~\ref{app:3N_1st}.
As in Refs.~\cite{Hebeler:2009iv,Hebeler:2010xb}, we use a smooth
regulator $f_\text{R}(p,q)= \exp \bigl[-((p^2+ 3 q^2/4)/
  \Lambda_\text{3N}^2)^{4}\bigr]$ with Jacobi momenta $p$ and $q$,
which is symmetric under exchange of any particles. The $c_D, c_E$
couplings have been fit in Ref.~\cite{Hebeler:2010xb} for given
$V_{{\rm low}\,k}$, $c_i$ couplings, and $\Lambda/\Lambda_\text{3N}$
to the $^3$H binding energy and the point charge radius of $^4$He.
This strategy has also been adopted to study exotic nuclei
(see, e.g., Refs.~\cite{Otsuka:2009cs,Holt:2010yb}) with recent
experimental highlights~\cite{Gallant:2012as,Wienholtz:2013nya}.

We consider the seven interaction sets given in Table~\ref{tab:sets},
where the $\Lambda/\Lambda_\text{3N}$ cutoffs and the $c_i$ couplings
are varied. This includes the consistent $c_i$'s of the N$^3$LO $500
\, {\rm MeV}$ NN potential of Ref.~\cite{Entem:2003ft} (sets~1-5), the
$c_i$'s from the N$^3$LO potentials of Ref.~\cite{Epelbaum:2004fk}
(set~6) and from the NN partial wave
analysis~\cite{Rentmeester:2003mf} (set~7). 
For the latter two $c_i$
sets (6 and 7), the $c_i$ couplings in the 3N force are not consistent
with the NN interaction. For the purpose of this work, we consider the
$c_i$ variation as a probe of the uncertainty from higher-order
long-range 3N forces (see Ref.~\cite{Hebeler:2009iv,Tews:2012fj,Kruger:2013kua}).
For the results, we will
take the energy range given by these interaction sets as a measure of
the theoretical uncertainty~\cite{Hebeler:2009iv, Hebeler:2010xb}. This probes the sensitivity to neglected
higher-order short-range couplings (from cutoff variation) and the
uncertainties in the long-range parts of 3N forces (from $c_i$
variation). To improve upon this, future calculations will include the
N$^3$LO 3N and 4N interactions following
Refs.~\cite{Tews:2012fj,Kruger:2013kua} and the consistent similarity
RG evolution of 3N interactions in momentum
space~\cite{Hebeler:2012pr,Hebeler:2013ri}.

\begin{table}[t]
\caption{Different sets of 3N couplings employed in the present 
calculations, taken from Ref.~\cite{Hebeler:2010xb}. The values
of the dimensionless $c_D$ and $c_E$ are fit to the
$^3$H binding energy $E_{^3\text{H}} = -8.482 \, \text{MeV}$ and 
the point charge radius of $^4$He $r_{^4\text{He}} = 1.464 \,
\text{fm}$ for the different NN/3N cutoffs and different $c_i$
couplings. $\Lambda/\Lambda_\text{3N}$ are in fm$^{-1}$
and the $c_i$ are in GeV$^{-1}$.\label{tab:sets}}
\begin{center}
\begin{tabular}{c|cc|ccc|cc}
\hline\hline
\, set \, & \, $\Lambda$ \, & \, $\Lambda_\text{3N}$ \, 
& \, $c_1$ \, & \, $c_3$ \, & \, $c_4$ \, 
& \, $c_D$ \, & \, $c_E$ \, \\
\hline
1 & 1.8 & 2.0 & $-0.8$ & $-3.2$ & 5.4 & $-1.621$ & $-0.143$ \\
2 & 2.0 & 2.0 & $-0.8$ & $-3.2$ & 5.4 & $-1.705$ & $-0.109$ \\
3 & 2.0 & 2.5 & $-0.8$ & $-3.2$ & 5.4 & $-0.230$ & $-0.538$ \\
4 & 2.2 & 2.0 & $-0.8$ & $-3.2$ & 5.4 & $-1.575$ & $-0.102$ \\
5 & 2.8 & 2.0 & $-0.8$ & $-3.2$ & 5.4 & $-1.463$ & $-0.029$ \\
6 & 2.0 & 2.0 & $-0.8$ & $-3.4$ & 3.4 & $-4.381$ & $-1.126$ \\
7 & 2.0 & 2.0 & $-0.8$ & $-4.8$ & 4.0 & $-2.632$ & $-0.677$ \\
\hline\hline
\end{tabular}
\end{center}
\end{table}
	
\subsection{Calculational strategy}

We focus on the calculation of asymmetric nuclear matter with small
proton fraction (for neutron-rich conditions). Our calculational
scheme relies on the result that neutron matter is perturbative for low
momentum interactions~\cite{Hebeler:2009iv}, which was also shown
recently for chiral EFT interactions with low
cutoffs~\cite{Tews:2012fj,Kruger:2013kua} and validated with Quantum
Monte Carlo~\cite{Gezerlis:2013ipa}. 
Note that even the largest NN cutoff interaction (set 5) has been shown to be 
perturbative in symmetric nuclear matter~\cite{Hebeler:2010xb}.
We include NN and 3N interactions
at the Hartree-Fock level and perturbative corrections to the energy
density $E/V$ from NN interactions at second order:
\begin{equation}
\frac{E_\text{NN}}{V} \approx \frac{E_\text{NN}^{(1)}}{V} + 
\frac{E_\text{NN}^{(2)}}{V} \hspace*{3mm} {\rm and} \hspace*{3mm}
\frac{E_\text{3N}}{V} \approx \frac{E_\text{3N}^{(1)}}{V} \,.
\label{eq:mbpt2}
\end{equation}
This was found to be a reliable approximation for neutron
matter~\cite{Hebeler:2009iv}. In particular, note that second-order
corrections involving 3N interactions have been shown to contribute
only at the hundred keV level in neutron matter, see Table~I of
Ref.~\cite{Hebeler:2009iv}.

Asymmetric nuclear matter is characterized by the neutron and proton
densities, $n_n$ and $n_p$, or equivalently by the proton fraction $x
= n_p/n$ and the density $n = n_p + n_n$. In addition, we recall that
for a given $x$, the proton and neutron Fermi momenta, $k_F^p$ and
$k_F^n$, and the density $n$ are related by $k_F^p = k_F^n
[x/(1-x)]^{1/3}$ and $n = (k_F^n)^3/[3 \pi^2 (1-x)]$.

We consider proton fractions $x \leqslant 0.15$. For such neutron-rich
conditions, the contributions involving two and three protons are
small, so that we approximate
\begin{equation}
\frac{E_{\text{NN}}}{V} \approx \frac{E_{nn}}{V} + \frac{E_{np}}{V}
\hspace*{3mm} {\rm and} \hspace*{3mm}
\frac{E_{\text{3N}}}{V} \approx \frac{E_{nnn}}{V} + \frac{E_{nnp}}{V} \,.
\label{eq:smallx}
\end{equation}
As a check, we have evaluated the $pp$, $ppn$ and $ppp$ contributions
at the Hartree-Fock and second-order NN level. As discussed in the
following, for the largest proton fraction considered ($x=0.15$),
these lead to energy contributions $[E_{pp}+E_{ppn}+E_{ppp}]/A = -0.2
\, {\rm MeV}$ at saturation density $n_0 = 0.16 \, {\rm fm}^{-3}$,
which are small compared to our uncertainty bands (see
Fig.~\ref{fig:energies}). We emphasize that closer to symmetric
nuclear matter, the inclusion of higher-order many-body contributions
will be important~\cite{Hebeler:2010xb}. Work is under way to include
these and to relax the approximation in the number of proton lines.

\subsection{First-order NN contribution}
\label{sec:NN_1st}

The NN Hartree-Fock contribution to the energy density is given by
\begin{align}
\frac{E_\text{NN}^{(1)}}{V} &= \frac{1}{2} \sum\limits_{T,M_T}
\int \frac{d\vec{k}}{(2\pi)^6} \left( \int d\vec{P} \ 
n_{\frac{\vec{P}}{2}+\vec{k}}^{\tau_1} \, n_{\frac{\vec{P}}{2}-\vec{k}}^{\tau_2} 
\right) \nonumber \\[1mm]
&\quad \times \sum_{S,M_S} \langle \vec{k} S M_S T M_T | \mathcal{A}_{12}	
V_\text{NN} |\vec{k} S M_S T M_T \rangle \,,
\label{eq:NN_1st_start}
\end{align}
where $\vec{k} = (\vec{k}_1-\vec{k}_2)/2$ and $\vec{P} =
\vec{k}_1+\vec{k}_2$ are the relative and center-of-mass momentum,
$n_{\bf k_i}^{\tau_i}$ are the Fermi distribution functions of species
$\tau_i=n, p$, and $S$, $T$ denote the two-body spin and isospin, with
projections $M_S, M_T$. For $M_T=0$, Eq.~(\ref{eq:NN_1st_start})
implies that $\tau_1=n$ and $\tau_2=p$. The energy involves a
spin-summed antisymmetrized matrix element of the NN interaction with
antisymmetrizer $\mathcal{A}_{12} = 1 - P_{12}$, where the
particle-exchange operator $P_{12} = P_{12}^{k} \, P_{12}^{\sigma} \,
P_{12}^{\tau}$ acts on momentum, spin, and isospin.

The integral over the center-of-mass momentum in
Eq.~\eqref{eq:NN_1st_start} can be performed separately, as the NN
interaction matrix element is independent of $\vec{P}$. The
integration results in a function $f^{M_T}\hspace{-0.5mm}(k)$, which
is given in Appendix~\ref{app:2N_1st}. Expanding the NN matrix element
in partial waves, the $M_S$ sum can be performed explicitly. This
gives for the NN Hartree-Fock energy density
\begin{align}
\frac{E_{\text{NN}}^{(1)}}{V} &= \frac{1}{8\pi^4} \int \limits_0^{\frac{k_F^n+k_F^p}{2}} dk \, k^2 \, \sum_{l,S,J} (2J+1) \nonumber \\ 
&\quad \times \bigg[ f^{nn}(k) \, \langle k| V_{l,l}^{J,S,M_T=-1}|k\rangle \left(1-(-1)^{l+S+1}\right) \nonumber \\ 
&\quad + f^{np}(k) \, \langle k| V_{l,l}^{J,S,M_T=0}|k\rangle \left(1-(-1)^{l+S}\right) \nonumber \\
&\quad + f^{np}(k) \, \langle k| V_{l,l}^{J,S,M_T=0}|k\rangle \left(1-(-1)^{l+S+1}\right) \bigg] \,,
\label{eq:NN_1st_contr}
\end{align}
where we have neglected the $pp$ contribution according to the
approximation~\eqref{eq:smallx} and $f^{M_T=0} \equiv f^{np}$. The
orbital and total angular momentum are labeled by $l$ and
$J$, respectively, and the factor $\left(1-(-1)^{l+S+T}\right)$ takes
into account the exchange term.

\subsection{Second-order NN contribution}
\label{sec:NN_2nd}

The second-order NN contribution to the energy density reads
\begin{align}
\frac{E^{(2)}_\text{NN}}{V} &= \frac{1}{4} \sum \limits_{S,M_S,M_{S'},T,M_T}  \int \frac{d\vec{k} \, d\vec{k}' \, d\vec{P}}{(2 \pi)^9} \nonumber \\[1mm]
& \quad \times \frac{n^{\tau_1}_{\frac{\vec{P}}{2}+\vec{k}} \, 
n^{\tau_2}_{\frac{\vec{P}}{2}-\vec{k}} \, (1-n^{\tau_3}_{\frac{\vec{P}}{2}+\vec{k}'}) \, 
(1-n^{\tau_4}_{\frac{\vec{P}}{2}-\vec{k}'})}{(k^2-k'^2)/m} \nonumber \\ 
& \quad \times \big| \bra{\vec{k} S M_S T M_T} \mathcal{A}_{12} V_\text{NN} \ket{\vec{k}' S M_S' T M_T} \big|^2 \,,
\label{eq:NN_2nd_start}
\end{align}
where $\vec{k'}=(\vec{k}_3-\vec{k}_4)/2$ and we use an averaged
nucleon mass $m=938.92$\,MeV. In addition, for $M_T=0$ also $\tau_3=n$
and $\tau_4=p$. Expanding the NN matrix elements in partial waves and
after spin sums, we have~\cite{Hebeler:2009iv}
\begin{align}
&\sum \limits_{S, M_S,M_S'} \big| \bra{\vec{k} S M_S T M_T} \mathcal{A}_{12} V_\text{NN} \ket{\vec{k}' S M_S' T M_T} \big|^2 \nonumber \\
&= \sum_{L,S} \sum \limits_{J,l,l'} \sum \limits_{\widetilde{J},\widetilde{l},\widetilde{l}'} P_L(\cos \theta_{\vec{k},\vec{k'}}) (4\pi)^2 \, i^{(l-l'+\widetilde{l}-\widetilde{l}')} \, (-1)^{\widetilde{l}+l'+L} \nonumber \\
&\quad \times \clebschG{l}{\widetilde{l}'}{L}{0}{0}{0} \clebschG{l'}{\widetilde{l}}{L}{0}{0}{0}\sqrt{(2l+1)(2l'+1)(2\widetilde{l}+1)(2\widetilde{l}'+1)} \nonumber \\
&\quad \times (2J+1)(2\widetilde{J}+1)
\begin{Bmatrix}
l & S & J \\
\widetilde{J}&L&\widetilde{l}'
\end{Bmatrix}
\begin{Bmatrix}
J & S & l' \\
\widetilde{l}&L&\widetilde{J}
\end{Bmatrix} \nonumber \\
&\quad \times \bra{k}V_{l',l}^{J, S, M_T}\ket{k'} \bra{k'}V_{\widetilde{l}',\widetilde{l}}^{\widetilde{J},S, M_T}\ket{k} \nonumber \\
&\quad \times \bigl(1-(-1)^{l+S+T}\bigr) \bigl(1-(-1)^{\widetilde{l}+S+T}\bigr) \,,
\label{eq:NN_2nd_pw}
\end{align}
with Legendre polynomial $P_L$, Clebsch-Gordan coefficients ${\mathcal
C}$, and $6J$-symbols. We consider only the $L=0$ contribution in
the partial-wave sum~\eqref{eq:NN_2nd_pw}, which is equivalent
to angle averaging. The spin-summed NN matrix elements are then angle
independent and the angular integrations over the Fermi distribution
functions in Eq.~\eqref{eq:NN_2nd_start} can be performed
analytically, leading to the function
\begin{align}
F^{M_T}\hspace{-0.5mm}(k,k',P) &= \int d\Omega_\vec{k} \int d\Omega_{\vec{k}'} 
\int d\Omega_\vec{P} \nonumber \\[1mm]
&\quad \times n^{\tau_1}_{\frac{\vec{P}}{2}+\vec{k}} \, 
n^{\tau_2}_{\frac{\vec{P}}{2}-\vec{k}} \, 
(1-n^{\tau_3}_{\frac{\vec{P}}{2}+\vec{k}'}) \, 
(1-n^{\tau_4}_{\frac{\vec{P}}{2}-\vec{k}'}) \,,
\end{align}
which is derived in detail in Appendix~\ref{app:2N_2nd}. Combining
this, we obtain for the second-order NN contribution to the energy
density
\begin{align}
\frac{E^{(2)}_\text{NN}}{V} &= \frac{1}{4} \frac{1}{(2 \pi)^9} 
\int\limits_0^{k_F^n+k_F^p} dP \, P^2 \int\limits_0^{\frac{k_F^n+k_F^p}{2}} 
dk \, k^2 \int \limits_0^\infty dk' \, k'^2 \nonumber \\
& \times \frac{m}{k^2-k'^2} \sum \limits_{S,M_S,M_{S'},T,M_T} 
F^{M_T}\hspace{-0.5mm}(k,k',P) \nonumber \\ 
& \times \big| \bra{\vec{k} S M_S T M_T} \mathcal{A}_{12} V_\text{NN} \ket{\vec{k}' S M_S' T M_T} \big|^2 \,,
\label{eq:NN_2nd_contr}
\end{align}
where the spin-isospin-summed matrix elements are given explicitly by
Eq.~(\ref{eq:2ndsum}), which neglects the $pp$ contributions,
multiplied by the appropriate phase-space functions
$F^{M_T}\hspace{-0.5mm}(k,k',P)$ in each channel.

\subsection{First-order 3N contribution}
\label{sec:3N_1st}

The 3N Hartree-Fock contribution to the energy density is given by
\begin{align}
\frac{E_\text{3N}^{(1)}}{V} &= \frac{1}{6} \, \text{Tr}_{\sigma_1,\tau_1} 
\text{Tr}_{\sigma_2,\tau_2} \text{Tr}_{\sigma_3,\tau_3} \int \frac{d \vec{k}_1 \,
d \vec{k}_2 \, d \vec{k}_3}{(2 \pi)^9} \nonumber \\[1mm]
& \quad \times n^{\tau_1}_{\vec{k}_1} \, n^{\tau_2}_{\vec{k}_2} \, 
n^{\tau_3}_{\vec{k}_3} \, f_\text{R}^2 \, \bra{123} \mathcal{A}_{123} \,		V_{\text{3N}} \ket{123} \,,
\label{eq:3N_1st_start}
\end{align}
where $i \equiv {\bf k}_i, \sigma_i, \tau_i$ is a short-hand notation
that includes all single-particle quantum numbers, $f_R$ is the
three-body regulator, and the three-body antisymmetrizer
$\mathcal{A}_{123}$ is
\begin{align}
\mathcal{A}_{123} &= (1 + P_{12} P_{23} + P_{13} P_{23}) (1 - P_{23}) \, 
\nonumber \\
&= 1-P_{12}-P_{13}-P_{23}+P_{12}P_{23}+P_{13}P_{23} \,.
\end{align}
In the present work, we only include the contributions involving two
or three neutrons due to the approximation~\eqref{eq:smallx}.
However, for isospin-symmetric interactions, the other contributions
follow simply from exchanging neutrons with protons.

The contribution from three neutrons to the energy density,
$E^{(1)}_{nnn}/V$ in Eq.~\eqref{eq:smallx}, has been derived in the
neutron matter calculation of Ref.~\cite{Hebeler:2009iv}. In this
case, the $c_4$ part of $V_c$, as well as the $V_D$ and $V_E$ terms
vanish (with the non-local regulator $f_R$) due to their isospin
structure ($c_4$), the Pauli principle ($V_E$) and the coupling of
pions to spin ($V_D$)~\cite{Hebeler:2009iv}. For the contributions
involving two neutrons and a proton, $E^{(1)}_{nnp}/V$, all parts of
the N$^2$LO 3N interactions enter.  Their derivation is discussed in
detail in Appendix~\ref{app:3N_1st}, where the final expressions for
the $V_c$, $V_D$ and $V_E$ parts are given by
Eqs.~\eqref{eq:3N_VC_contr}, \eqref{eq:3N_VD_contr} and
\eqref{eq:3N_VE_contr}. In summary, the 3N Hartree-Fock energy density
neglecting the contributions from two and more proton lines is given
by
\begin{equation} 
\frac{E_{3N}^{(1)}}{V} = \frac{E_{V_c}^{(1)}}{V} \Bigg|_{nnn} 
+ 3 \left( \frac{E_{V_c}^{(1)}}{V} + \frac{E_{V_D}^{(1)}}{V} + 
\frac{E_{V_E}^{(1)}}{V} \right) \Bigg|_{nnp} \,.
\label{eq:3N_1st_contr}
\end{equation}

\section{Results}
\label{sec:results}

\subsection{Energy of asymmetric nuclear matter}
\label{sec:energy}

\begin{figure*}[t]
\begin{center}
\vspace*{-2mm}
\includegraphics[width=0.875\textwidth,clip=]{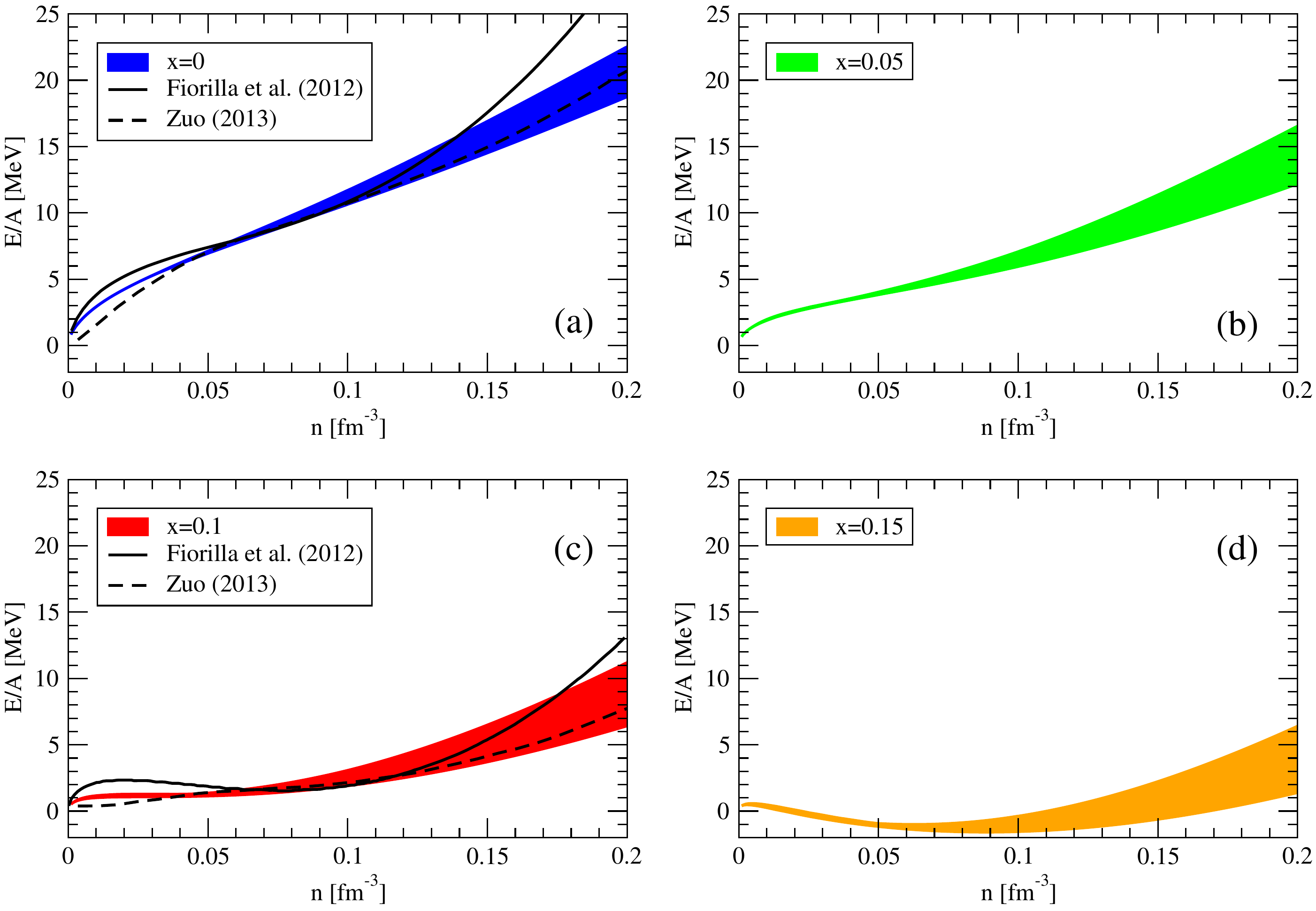}
\vspace*{-4mm}
\end{center}
\caption{(Color online) Energy per particle $E/A$ of pure neutron matter 
($x=0$) and asymmetric nuclear matter for three different proton fractions
$x=0.05, 0.1$, and $0.15$ as a function of density $n$. The bands
estimate the uncertainty of our calculations (see text for details).
Where available, we compare our results to the Brueckner-Hartree-Fock
energies of Ref.~\cite{Zuo:2012sa} (Zuo) and to the energies obtained
from in-medium chiral perturbation theory (Fiorilla et 
al.)~\cite{Fiorilla:2011sr}.\label{fig:energies}}
\end{figure*}

We calculate the energy of asymmetric nuclear matter by evaluating
Eqs.~\eqref{eq:NN_1st_contr},~\eqref{eq:NN_2nd_contr}
and~\eqref{eq:3N_1st_contr} for densities $n \leqslant 0.2 \,
\text{fm}^{-3}$ and proton fractions $x \leqslant 0.15$. Our results
for the energy per particle $E/A$ are presented in
Fig.~\ref{fig:energies} for pure neutron matter ($x=0$) and for three
different proton fractions ($x=0.05, 0.1$, and $0.15$). As discussed
in Sect.~\ref{sec:H}, we perform calculations for a range of cutoffs
and $c_i$ couplings, which gives an estimate of the theoretical
uncertainty. This range is larger than the one from approximations in
the many-body calculation~\cite{Hebeler:2009iv,Hebeler:2010xb}. In
Fig.~\ref{fig:energies} and in the following, this uncertainty
estimate is presented as energy bands. We emphasize that 3N forces are
fit only to light nuclei and no parameters are adjusted to empirical
nuclear matter properties.

\begin{figure*}[t]
\begin{center}
\vspace*{-2mm}
\includegraphics[clip=,width=0.875\textwidth]{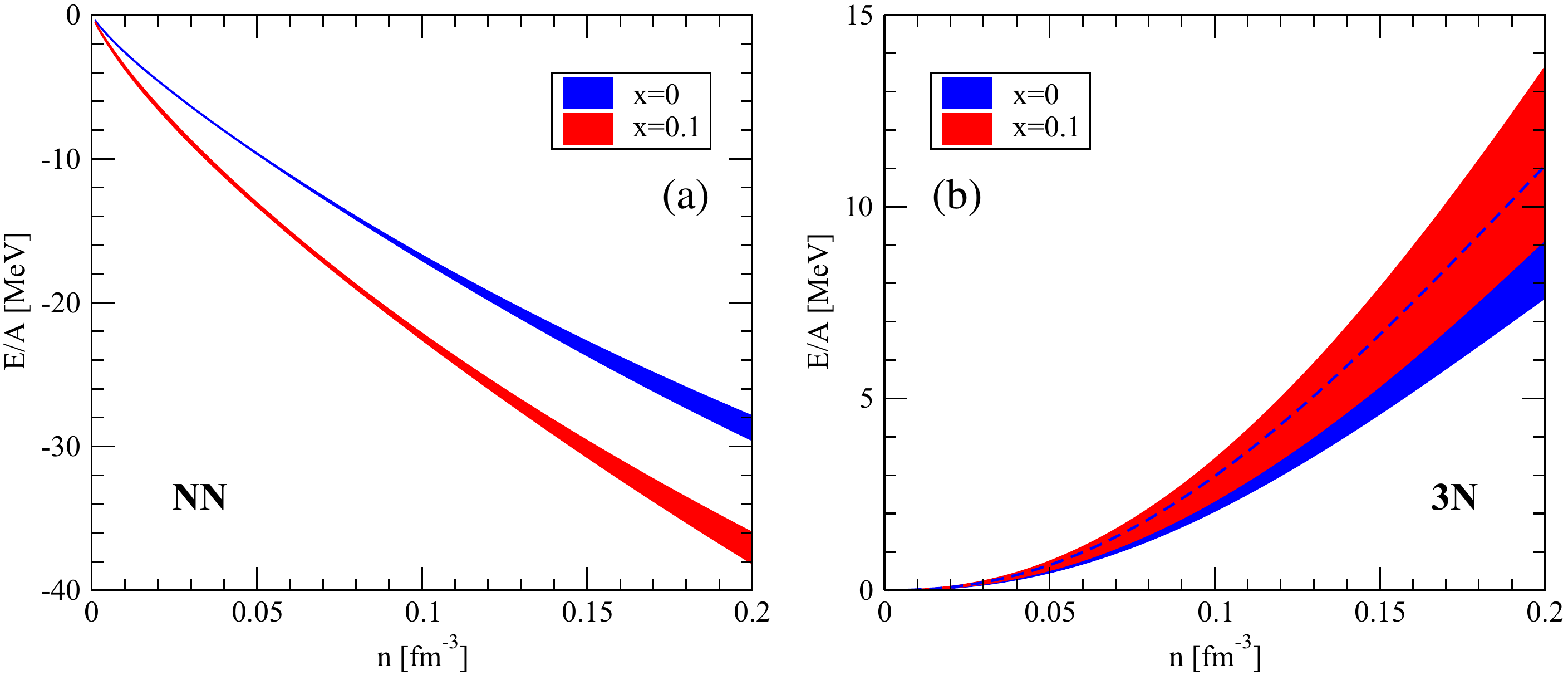}
\vspace*{-4mm}
\end{center}
\caption{(Color online) Interaction energy per particle from NN (left
panel) and 3N (right panel) contributions for pure neutron matter (blue)
and asymmetric nuclear matter with proton fraction $x=0.1$ (red bands)
as a function of density.\label{fig:energies_NN_3N}}
\end{figure*}

The energy per particle in neutron matter has been benchmarked with
the values reported in Ref.~\cite{Hebeler:2009iv}, with excellent
agreement. For two proton fractions ($x=0$ and $0.1$), we compare our
energies to explicit calculations of asymmetric nuclear matter. The
Brueckner-Hartree-Fock results of Ref.~\cite{Zuo:2012sa} (Zuo) are
based on the Argonne $v_{18}$ supplemented by phenomenological 3N
forces of Ref.~\cite{Grange:1989nx}. While they exhibit an unusual
behavior at low densities, they lie within our bands for densities $n
\gtrsim 0.05 \, \text{fm}^{-3}$. In addition, we compare with the
results obtained from in-medium chiral perturbation theory (Fiorilla
et al.)~\cite{Fiorilla:2011sr}, which differ in their density
dependence compared to our ab-initio calculations. This could be
due to the approximation to the leading-order contact interactions
in Ref.~\cite{Fiorilla:2011sr}.

The interaction energies from NN and 3N contributions are shown
separately in Fig.~\ref{fig:energies_NN_3N} for two different proton
fractions ($x=0$ and $0.1$). We observe that the uncertainties from 3N
forces dominate. This is consistent with the results for neutron
matter~\cite{Hebeler:2009iv} and can be improved by going to higher
order in chiral EFT interactions and in the many-body calculation.

In order to assess an error estimate of our approximation, we have calculated 
the contributions involving two or more proton lines that are neglected in 
Eq.~\eqref{eq:smallx}. 
For the different proton fractions at
saturation density, we compare the central energy from the seven
interaction sets of Table~\ref{tab:sets} evaluated at the same many-body level as
Eq.~\eqref{eq:mbpt2}. 
For $x = 0.05$,
$0.1$, and $0.15$, we obtain $E_{pp}/A =-0.2 \, \text{MeV} \: (0.4
\%), -0.4 \, \text{MeV} \: (1.3 \%)$, and $-0.9 \, \text{MeV} \: (2.4
\%)$, where the percentage number in parenthesis is relative to the NN
interaction energy. Similarly for the 3N contributions,
$[E_{ppn}+E_{ppp}]/A = 0.1 \, \text{MeV} \: (1.0 \%), 0.3 \,
\text{MeV} \: (3.3 \%)$, and $0.6 \, \text{MeV} \: (7.2 \%)$, where
the percentage number is relative to $[E_{nnn}+E_{nnp}]/A$. This shows
that the neglected contributions from two and more proton lines are
small. Furthermore, the NN and 3N contributions are opposite and to a
large extent cancel in the total energy per particle. This confirms
that the approximation~\eqref{eq:smallx} works well for the
neutron-rich conditions considered in this work.
However, when we compare to constraints for the symmetry energy
based on experiment around symmetric nuclei (see Fig.~\ref{fig:Esym_vs_n}), we have
decided to include the small contributions from two or more proton
lines. The corresponding changes of the symmetry energy are smaller
than the theoretical uncertainties.

\subsection{Quadratic expansion and symmetry energy} 
\label{sec:parabolic}

The technical difficulties of asymmetric matter calculations have
triggered approximate or phenomenological expansions for the nuclear
equation of state. Starting from the saturation point of symmetric
matter, the quadratic expansion expresses the energy of asymmetric
matter in terms of the asymmetry parameter $\beta~=~(n_n~-~n_p)/n = 1-2x$
as
\begin{equation}
\frac{E(n, \beta)}{A}= \frac{E(n,\beta=0)}{A} + S_v(n) \, 
\beta^2 + \mathcal{O}(\beta^4) \,,
\label{eq:parabolic}
\end{equation}
where $S_v$ is the symmetry energy. Provided that the equation of
state of symmetric matter is known, $S_v$ is the only input needed to
extrapolate to asymmetric matter at order $\beta^2$.  Originally
designed for small values of $\beta$, the quadratic expansion has
proven to be successful over a large range of asymmetries. Microscopic
calculations have validated the $\beta^2$ truncation, with only small
deviations away from symmetric
matter~\cite{Bombaci:1991zz,Zuo:2002sg}.

\begin{figure}[t]
\begin{center}
\includegraphics[clip=,width=0.975\columnwidth]{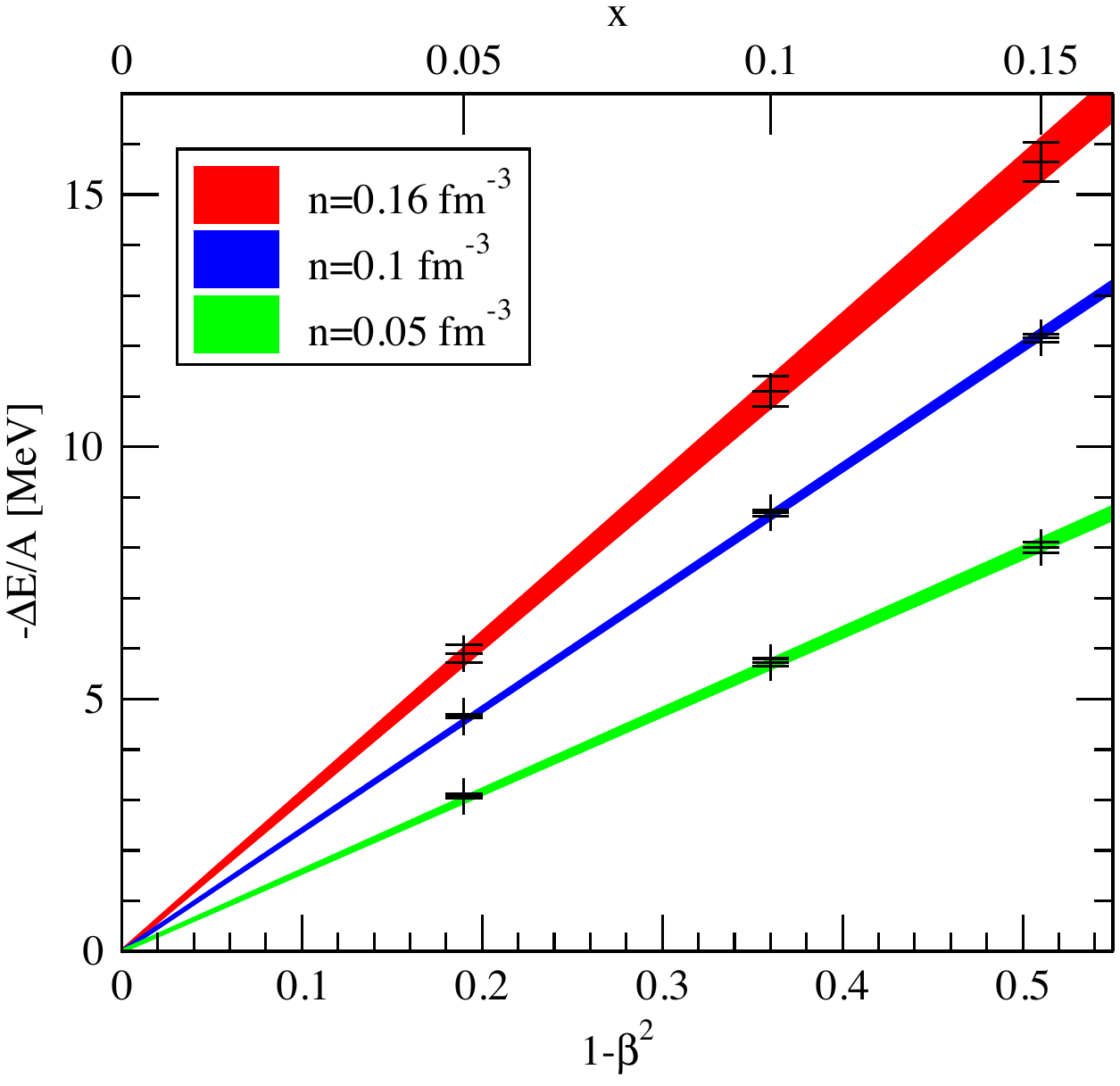}
\end{center}
\caption{(Color online) Energy per particle relative to pure neutron
matter $-\Delta E/A$ as a function of $(1-\beta^2)$ for three different
densities; the upper axis gives the proton fraction $x$. The points
correspond to our calculations, with error bars reflecting the
uncertainty bands of Fig.~\ref{fig:energies}. The colored bands are
linear fits to the points with the corresponding
errors.\label{fig:beta_plot}}
\end{figure}

We use our ab-initio calculations to test the quadratic expansion
for neutron-rich conditions. To this end, we define the energy
difference to pure neutron matter $\Delta E$:
\begin{equation}
\frac{\Delta E(n,x)}{A} = \frac{E(n, x)}{A} - \frac{E(n,x=0)}{A} \,.
\end{equation}
In terms of $\Delta E$, the quadratic approximation~\eqref{eq:parabolic}
reads
\begin{equation}
-\frac{\Delta E(n,\beta)}{A} = \frac{E(n,\beta=1)}{A} - 
\frac{E(n,\beta)}{A} = E_\text{sym}(n) \, (1-\beta^2) \,,
\label{eq:deltaE_fact}
\end{equation}
where $E_\text{sym}$ coincides with the symmetry energy $S_v$, if
$\mathcal{O}(\beta^4)$ terms vanish. Equation~\eqref{eq:deltaE_fact}
allows us to extract $E_\text{sym}$ for a given density and to verify
the linearity in $(1-\beta^2)$. In Fig.~\ref{fig:beta_plot}, we show
our results for $-\Delta E/A$ as a function of $(1-\beta^2)$ for three
representative densities.  For each value of $\beta$ (or $x$), the
vertical error bars reflect the energy range in
Fig.~\ref{fig:energies}. The colored bands in Fig.~\ref{fig:beta_plot}
are linear fits to the points with the corresponding errors. This
demonstrates that the quadratic expansion is a very good approximation
even for neutron-rich conditions.

\begin{figure}[t]
\begin{center}
\includegraphics[width=0.975\columnwidth,clip=]{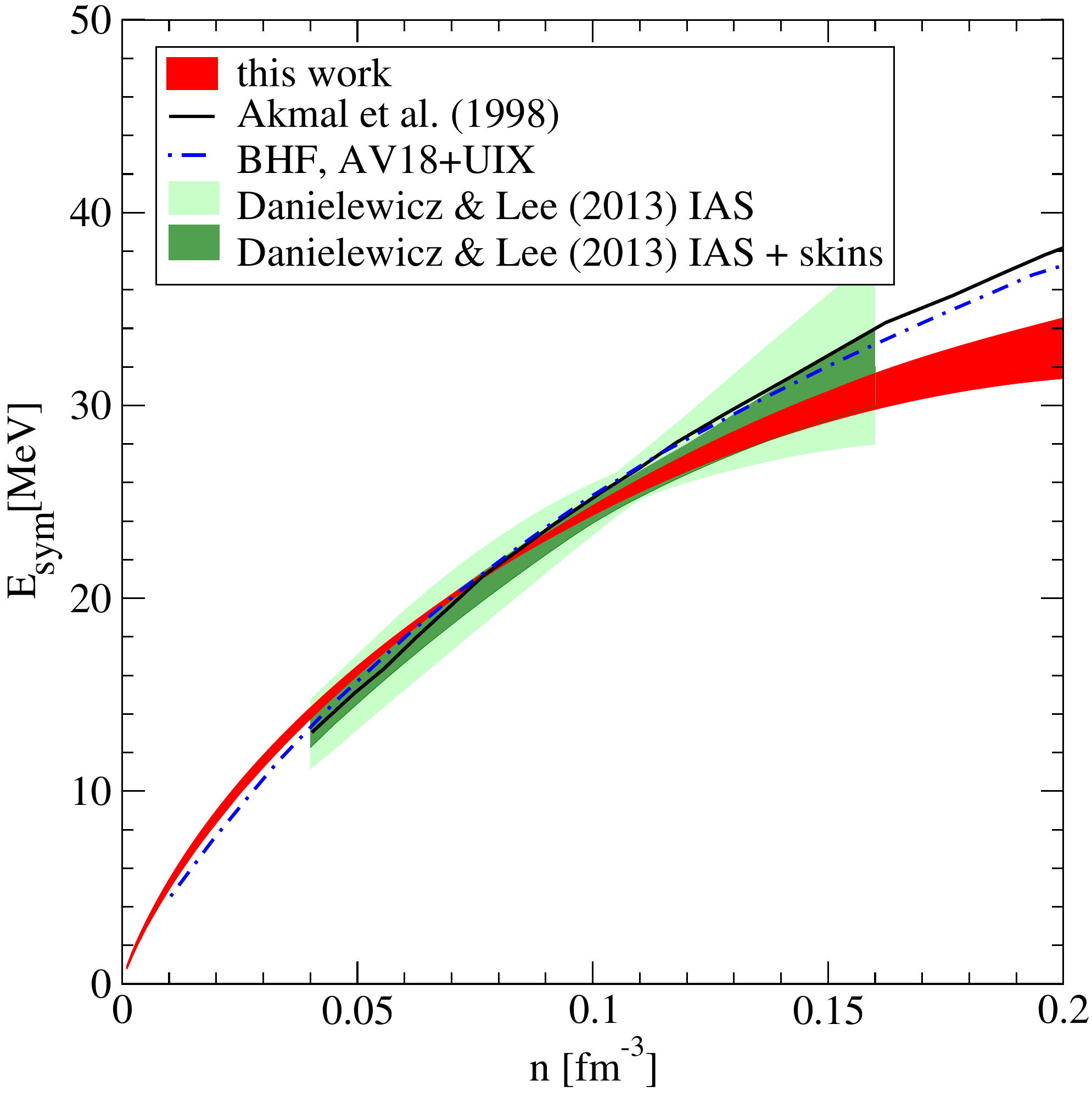}
\end{center}
\caption{(Color online) $E_\text{sym}$ as a function of density obtained
from our ab-initio calculations as in Fig.~\ref{fig:beta_plot}, 
including the small contributions from two or more proton lines.
In comparison, we give $E_\text{sym}$ obtained from microscopic
calculations performed with a variational approach (Akmal et
al.~(1998))~\cite{Akmal98} and at the Brueckner-Hartree-Fock level
(BHF)~\cite{Taranto13} based on the Argonne $v_{18}$ NN and Urbana UIX
3N potentials (with parameters adjusted to the empirical saturation
point). The band over the density range $n=0.04-0.16 \, {\rm fm}^{-3}$
is based on a recent analysis of isobaric analog states (IAS) and
including the constraints from neutron skins (IAS +
skins)~\cite{Danielewicz13}.\label{fig:Esym_vs_n}}
\end{figure}

\begin{table}[t]
\caption{$E_\text{sym}$ and corresponding uncertainties extracted
from the linear fits of Fig.~\ref{fig:beta_plot} for the three
densities.\label{tab:esym}}
\begin{center}
\begin{tabular}{c c}
\hline\hline
\, $n$ $[\text{fm}^{-3}]$ \, & \, $E_\text{sym}$ [MeV] \, \\
\hline
$0.05$ & $15.8 \pm 0.2$ \\
$0.10$ & $24.0 \pm 0.2$ \\
$0.16$ & $30.8 \pm 0.8$ \\
\hline\hline
\end{tabular}
\end{center}
\end{table}

\begin{figure*}[t]
\begin{center}
\includegraphics[width=0.975\textwidth]{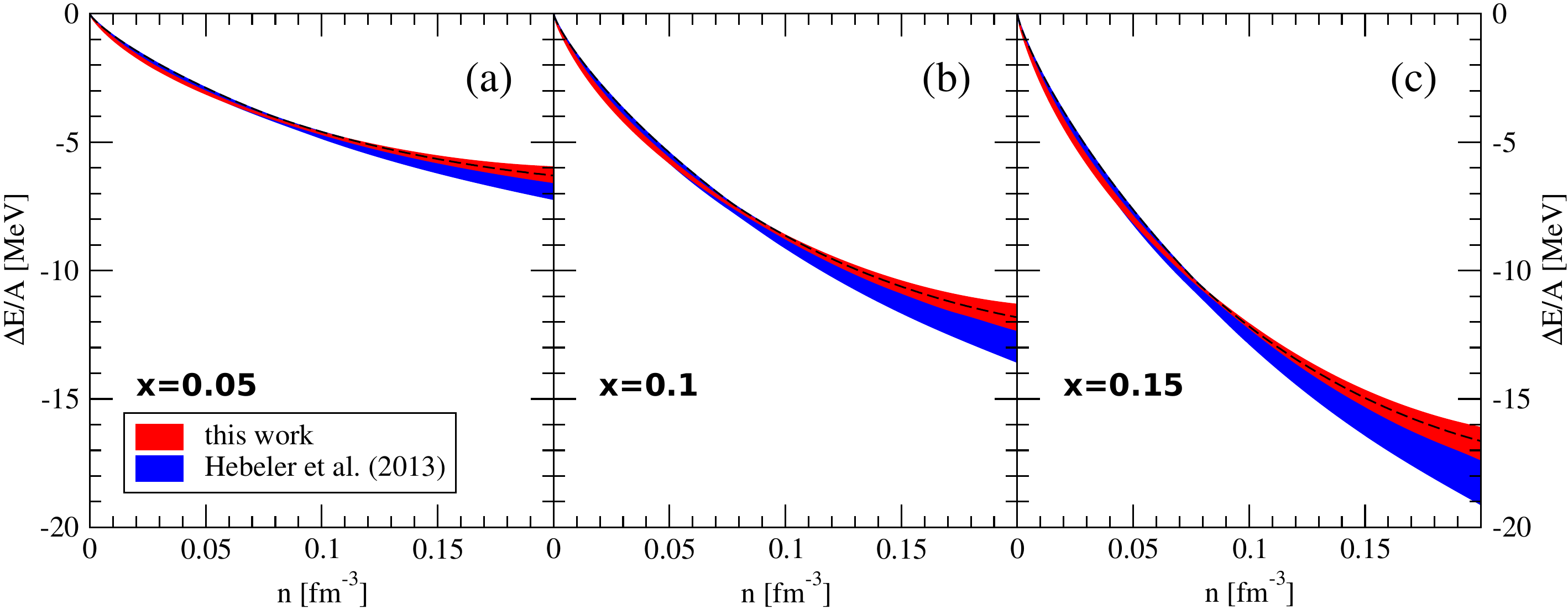}
\end{center}
\caption{(Color online) Energy per particle $\Delta E/A$ relative to pure
neutron matter as a function of density for three different proton
fractions $x=0.05, 0.1$, and $0.15$. The results of our calculations
(``this work'', red bands) are compared with the empirical
parametrization~\eqref{eq:expansion} used in
Ref.~\cite{Hebeler:2013nz} to extrapolate from pure neutron matter to
neutron-rich matter (Hebeler et al.~(2013), blue
bands).\label{fig:delta_energies}}
\end{figure*}

From the slope of the linear fits in Fig.~\ref{fig:beta_plot} one can
extract $E_\text{sym}$ for a given density. The resulting values for
the three representative densities are given in Table~\ref{tab:esym}.
At saturation density, we find $E_\text{sym} = 30.8 \pm 0.8 \,
\text{MeV}$. Note that with the inclusion of the contributions from
two or more proton lines, neglected in Eq.~\eqref{eq:smallx},
$E_\text{sym}$ slightly increases to $31.2 \pm 1.0 \, \text{MeV}$.
The uncertainty range is smaller than extracting $E_\text{sym}$ from
neutron matter calculations and the empirical saturation point (see
Refs.~\cite{Hebeler:2009iv,Hebeler:2010jx,Hebeler:2013nz}). This is
due to the explicit information from asymmetric matter results.

Figure~\ref{fig:Esym_vs_n} shows $E_\text{sym}$ as a function of
density extracted from our asymmetric matter calculations as in
Fig.~\ref{fig:beta_plot}. The $E_\text{sym}$ band is due to the
theoretical uncertainty of our calculations for the energy. In this
case, we have included the small contributions from two and more hole
lines discussed above. Our results are compared in
Fig.~\ref{fig:Esym_vs_n} with constraints from a recent analysis of
isobaric analog states (IAS) and including the constraints from
neutron skins (IAS + skins)~\cite{Danielewicz13}, showing a remarkable
agreement over the entire density range.  In addition, we show
$E_\text{sym}$ obtained from microscopic calculations performed with a
variational approach (Akmal et al. (1998))~\cite{Akmal98} and at the
Brueckner-Hartree-Fock level (BHF)~\cite{Taranto13}. Both calculations
are based on the Argonne $v_{18}$ NN and Urbana UIX 3N potentials
(with different parameters adjusted to the empirical saturation
point), but derive $E_\text{sym}$ from symmetric and pure neutron
matter using the quadratic expansion~\eqref{eq:parabolic}. These
results are compatible with our $E_\text{sym}$ band at low and
intermediate densities but predict a somewhat stiffer $E_\text{sym}$
for $n \gtrsim n_0$. We attribute these differences to the
phenomenological 3N forces used.

\subsection{Empirical parametrization}
\label{sec:expansion}

In order to extend ab-initio calculations of neutron matter to
asymmetric matter for astrophysical applications,
Ref.~\cite{Hebeler:2013nz} used an empirical parametrization that
represents an expansion in Fermi momentum with kinetic energies plus
interaction energies that follow the quadratic expansion with $x (1-x) =
(1 - \beta^2)/4$:
\begin{align}
\frac{E(\overline{n},x)}{A} &= T_0 \bigg[ \frac{3}{5} \left( 
x^{5/3}+(1-x)^{5/3} \right)(2\overline{n})^{2/3} 
\nonumber \\
&\quad - \left( (2 \alpha-4 \alpha_L) x (1-x)+\alpha_L \right) \overline{n}
\nonumber \\
&\quad + \left( (2 \eta-4 \eta_L) x (1-x)+\eta_L \right) \overline{n}^{4/3} 
\bigg] \,,
\label{eq:expansion}
\end{align}
where $\overline{n}=n/n_0$ denotes the density in units of saturation
density and $T_0=(3 \pi^2 n_0/2)^{2/3}/(2m) = 36.84 \, \text{MeV}$ is
the Fermi energy at $n_0$. The parameters $\alpha$, $\eta$, $\alpha_L$,
and $\eta_L$ are determined from fits to neutron-matter calculations
($\alpha_L, \eta_L$) and to the empirical saturation point of
symmetric matter. The latter gives $\alpha = 5.87, \eta = 3.81$. The
uncertainty range of $\alpha_L, \eta_L$ obtained from neutron-matter
calculations is shown in Fig.~4 of Ref.~\cite{Hebeler:2013nz}.

\begin{figure}[t]
\begin{center}
\includegraphics[clip=,width=0.95\columnwidth]{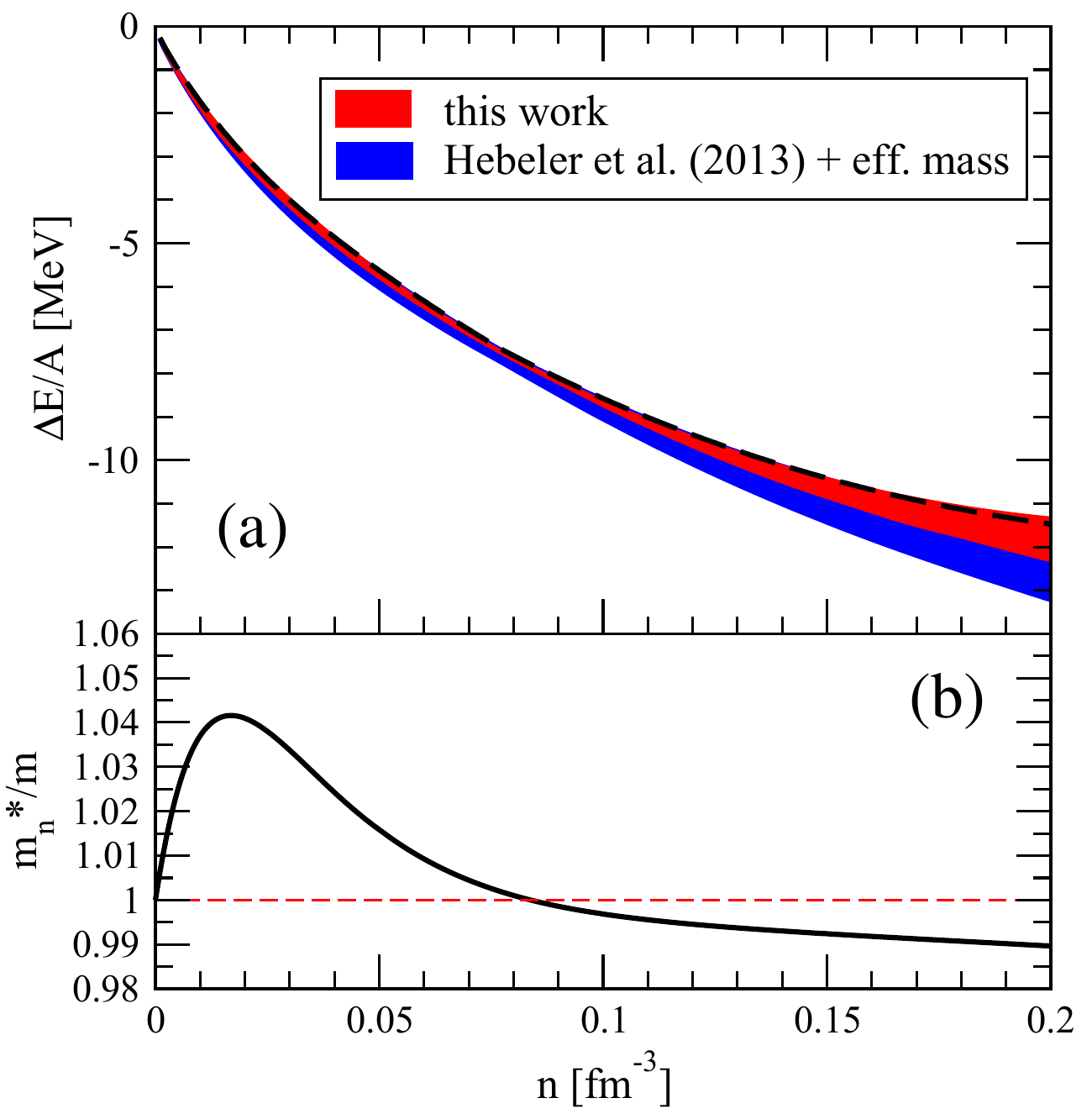}
\end{center}
\caption{(Color online) Upper panel:~Same as Fig.~\ref{fig:delta_energies} for a proton
fraction $x=0.1$ but with the modified kinetic term~\eqref{eq:Kmstar}
in the empirical parametrization~\eqref{eq:expansion}. Lower 
panel:~Neutron effective mass $m_n^*/m$ as a function of density 
obtained by fitting to $\Delta E/A$ of the upper panel (see text for
details).\label{fig:delta_energies_eff}}
\end{figure}

We use our ab-initio calculations to benchmark the empirical
parametrization~\eqref{eq:expansion} for asymmetric matter. The
comparison is shown in Fig.~\ref{fig:delta_energies} for the energy
difference to neutron matter $\Delta E/A$ as a function of density for
three different proton fractions. Remarkably, our results based on
nuclear forces fit only to few-body data agree within uncertainties
with the empirical parametrization~\eqref{eq:expansion} used in
Ref.~\cite{Hebeler:2013nz} to extrapolate from pure neutron matter to
neutron-rich matter. We observe only a slight difference in the
density dependence, with the empirical parameterization of Hebeler et
al.~\cite{Hebeler:2013nz} underestimating (overestimating) our band at
lower (higher) densities.

We investigate whether the small discrepancy could be due to a neutron
effective mass $m_n^*$ in the empirical expansion. To this end, we
replace the kinetic part in Eq.~\eqref{eq:expansion} by
\begin{equation}
T_0 \left[ \frac{3}{5} \left( x^{5/3} + \frac{m}{m^*_n}
(1-x)^{5/3} \right) (2\overline{n})^{2/3} \right] \,,
\label{eq:Kmstar}
\end{equation}
while the terms proportional to $\overline{n}$ and
$\overline{n}^{4/3}$ remain unchanged. For each proton fraction, we
fit a density-dependent neutron effective mass $m^*_n/m$ such that
the difference between (the upper bands of) our microscopic
calculation and the empirical parametrization with the modified
kinetic term~\eqref{eq:Kmstar} is minimized. The values and ranges for
$\alpha$, $\eta$, $\alpha_L$, and $\eta_L$ are kept the same. In
Fig.~\ref{fig:delta_energies_eff}, we show the resulting $m^*_n/m$
(lower panel) and the improved empirical parametrization (upper panel)
for a representative proton fraction $x=0.1$. With the introduction of
a weakly density-dependent neutron effective mass, the empirical
parametrization agrees excellently with our ab-initio results.
Moreover, the behavior of $m^*_n/m$ with a small increase at low
densities and a decreasing effective mass with increasing density is
in line with the expectations from microscopic
calculations~\cite{Schwenk:2002fq}.

Finally, we discuss the possible factorization of the dependence on
density and asymmetry in the energy of asymmetric nuclear matter. From
the three panels in Fig.~\ref{fig:delta_energies}, one notices that
increasing the proton fraction $x$ approximately results in an overall
rescaling of the density dependence of $\Delta E / A$. This rescaling
suggests a factorization of the dependence on $x$ and on the density:
$\Delta E/A (n,x) = \Psi(x) \Phi(n)$.  Such a factorization is
explicit in the quadratic expansion, where $\Psi(x)=x (1-x) = (1 -
\beta^2)/4$ and $\Phi(n) = -4 E_\text{sym}(n)$, see
Eq.~\eqref{eq:deltaE_fact}. Assuming the same $\Psi(x)=x (1-x)$, we
have checked whether a similar result holds for the empirical
parametrization~\eqref{eq:expansion}. In this case, our ab-initio
results for $\Delta E/A$ are approximately reproduced for $x \leqslant 0.15$ by
\begin{equation}
\Phi(n) = T_0 \Bigl[ -0.92 \, (2\overline{n})^{2/3} - (2 \alpha-4 \alpha_L)
\, \overline{n} + (2 \eta-4 \eta_L) \, \overline{n}^{4/3} \Bigr] .
\end{equation}
Using a central value of $\alpha_L = 1.33$ and $\eta_L = 0.88$ gives
an $E_\text{sym}(n) = - \Phi(n)/4$ that is very similar to our
ab-initio results in Fig.~\ref{fig:Esym_vs_n} and also lies within the
experimental constraints from IAS and neutron skins~\cite{Danielewicz13}.

\section{Conclusions} 
\label{sec:conclusion}

We have carried out the first calculations of asymmetric nuclear
matter with NN and 3N interactions based on chiral EFT. The phase
space due to the different neutron and proton Fermi seas was handled
without approximations. Focusing on neutron-rich conditions, we have
presented results for the energy of asymmetric matter for different
proton fractions (Fig.~\ref{fig:energies}), including estimates of the
theoretical uncertainty. As shown for neutron matter in
Ref.~\cite{Hebeler:2009iv}, the energy range is dominated by the
uncertainty in 3N forces (Fig.~\ref{fig:energies_NN_3N}).

We have used our ab-initio results to test the quadratic expansion
around symmetric matter with the symmetry energy term. The comparison
(Fig.~\ref{fig:beta_plot}) demonstrates that the quadratic
approximation works very well even for neutron-rich conditions. In
contrast to other calculations, our results are based on 3N forces fit
only to light nuclei, without adjustments to empirical nuclear matter
properties. Therefore, it is remarkable that the symmetry energy
extracted from our ab-initio calculations (Fig.~\ref{fig:Esym_vs_n})
is in very good agreement with empirical constraints from IAS and
neutron skins~\cite{Danielewicz13}. Moreover, compared to extracting
the symmetry energy from neutron-matter calculations and the empirical
saturation point, the symmetry-energy uncertainty is reduced due to
the explicit information from asymmetric matter.

Finally, we have studied an empirical parametrization of the energy
that represents an expansion in Fermi momentum with kinetic energies
plus interaction energies that are quadratic in the asymmetry. This
was used in Ref.~\cite{Hebeler:2013nz} to extend ab-initio
calculations of neutron matter to asymmetric matter for astrophysical
applications. Our asymmetric matter results are in remarkable
agreement with this empirical parametrization
(Fig.~\ref{fig:delta_energies}). This finding is very useful for
describing neutron-rich conditions in astrophysics, for neutron star
structure~\cite{Hebeler:2010jx,Hebeler:2013nz} and neutron star
mergers~\cite{Bauswein:2012ya}, and for developing new equations of
state for core-collapse supernovae.

The present calculations represent the first step of systematic
predictions of asymmetric nuclear matter including theoretical
uncertainties. This is very important in light of many astrophysical
applications. In the present work, we have limited our calculations to
neutron-rich conditions with $x \leqslant 0.15$. Future work includes
larger proton fractions, improvements in the many-body calculation,
and the inclusion of higher-order interactions in chiral EFT. These
are all possible due to recent developments~\cite{Gezerlis:2013ipa,%
Tews:2012fj,Kruger:2013kua,Hebeler:2012pr,Hebeler:2013ri}. It is
exciting that even at the current level, neutron-rich matter can be
reliably calculated and the results provide important input for
astrophysics. With the future improvements outlined above, we will
then be able to narrow the energy bands further.

\section*{Acknowledgements} 

We thank K.~Hebeler and I.~Tews for useful comments and
discussions. This work was supported by the Helmholtz Alliance Program
of the Helmholtz Association, contract HA216/EMMI "Extremes of Density
and Temperature: Cosmic Matter in the Laboratory", the DFG through
Grant SFB 634, and the ERC Grant No.~307986 STRONGINT.

\appendix

\section{Angular integrations and \\
partial-wave decomposition of NN contributions} 

\subsection{First-order NN contribution}
\label{app:2N_1st}

We first consider the NN contribution to the Hartree-Fock
energy~\eqref{eq:NN_1st_start}. The integral over the total momentum
of the nucleon pair can be performed separately, as the interaction is
independent of $\vec{P}$.  Taking the direction of $\vec{k}$ along the
$z$ axis, the integration yields a function of $k$,
\begin{equation}
f^{np}(k) = \int d\vec{P} \ n^n_{\frac{\vec{P}}{2}+\vec{k}} \, 
n^p_{\frac{\vec{P}}{2}-\vec{k}} \,,
\label{eq:Pintegral}
\end{equation}
where we consider the general case of different Fermi seas. The case
of two neutrons/protons is then easily obtained.

\begin{figure}[t]
\centering
\subfloat[$\,$ Case (1.1): $0 \leqslant k\leqslant \frac{k_F^n-k_F^p}{2} \,.$]{\includegraphics[width=0.35\textwidth,clip=]{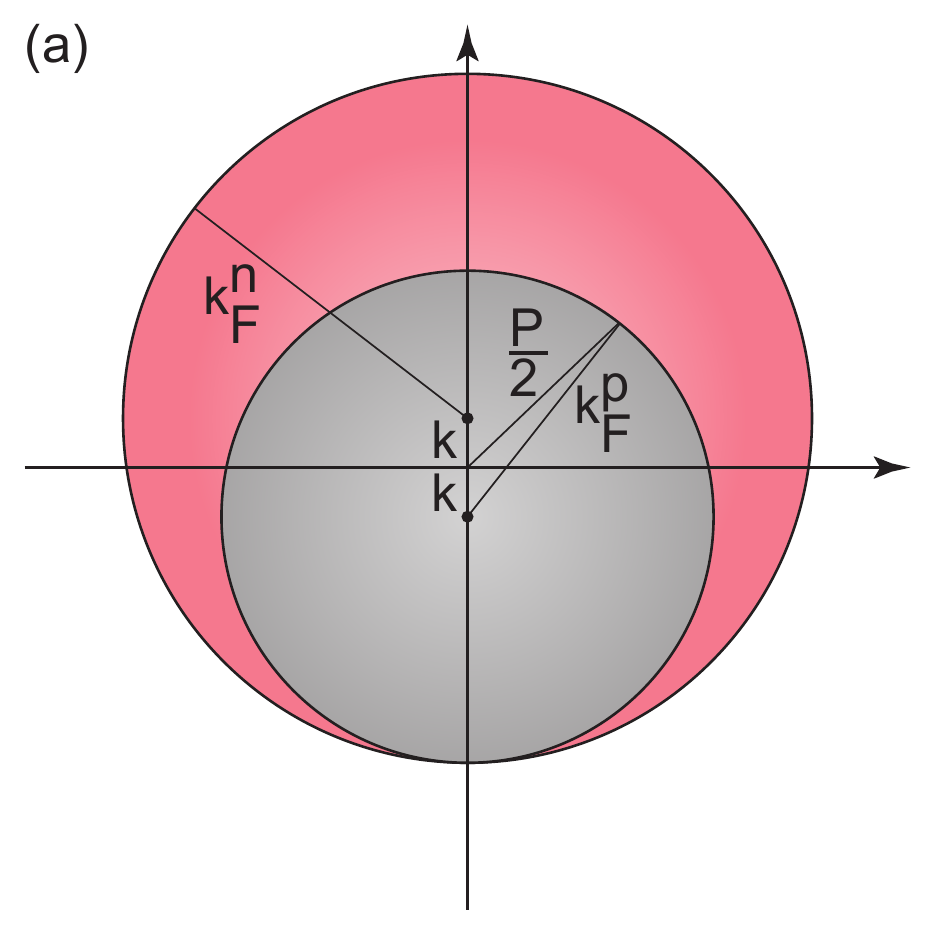}} \\
\subfloat[$\,$ Case (1.2): $\frac{k_F^n-k_F^p}{2} \leqslant k\leqslant \frac{k_F^n+k_F^p}{2} \,.$]{\includegraphics[width=0.35\textwidth,clip=]{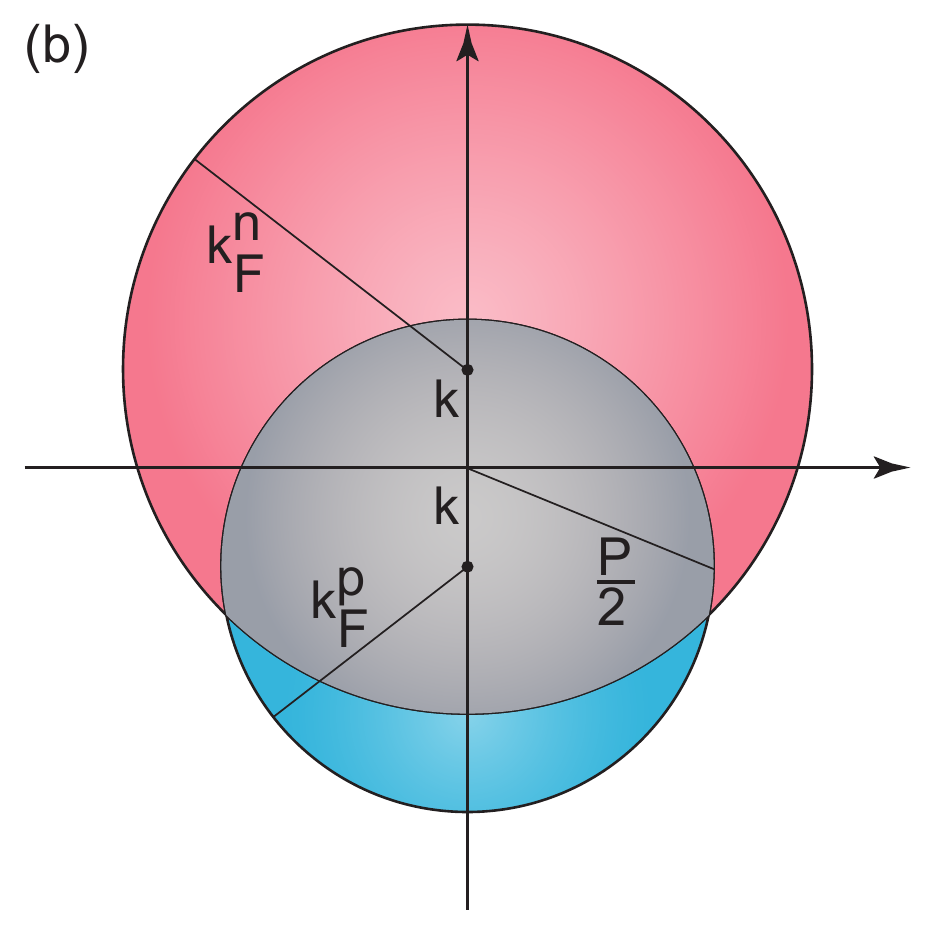}}
\caption{(Color online) Different regions contributing to the 
integral~\eqref{eq:Pintegral}. As discussed in the text, there are three
possible cases. The two non-vanishing ones are shown: the neutron
(red) and proton (blue) Fermi seas overlap totally~(a) or
partially~(b). Only the overlap (grey) contributes to the
integral.\label{fig:P_Integration_2_cases}}
\end{figure}

The two Fermi distribution functions are equivalent to two spheres in
momentum space, displaced by $\pm \vec{k}$ relative to the origin.
Assuming $k_F^n \geqslant k_F^p$, there are three possible
configurations depending on the value of $k$. The Fermi seas overlap
partially, totally, or they do not overlap:
\begin{itemize}
\item[(1.1)]  $\displaystyle \quad 0 \leqslant k \leqslant 
\frac{k_F^n-k_F^p}{2}$ \,,
\item[(1.2)]  $\displaystyle \quad \frac{k_F^n-k_F^p}{2} \leqslant k 
\leqslant \frac{k_F^n+k_F^p}{2}$ \,,
\item[(1.3)]  $\displaystyle  \quad k \geqslant \frac{k_F^n+k_F^p}{2}$ \,.
\end{itemize}
The first two cases are shown in
Fig.~\ref{fig:P_Integration_2_cases}. The case~(1.3) is trivial
because the integral vanishes. Here and in the following section, we
only give the non-vanishing cases. The angular integration yields
\begin{equation}
f^{np}(k) =
\begin{cases}
\frac{32 \pi}{3} (k_F^p)^3 & \text{for case (1.1)\,,} \\[1mm]
\frac{\pi}{3k} (-2 k+k_F^n+k_F^p)^2 \\
\times \bigl[ 4 k^2+4 k (k_F^n+k_F^p) \\
\quad -3 (k_F^n-k_F^p)^2 \bigr] & \text{for case (1.2)\,.} 
\end{cases}
\end{equation}
Case~(1.1) is simply 8 (from $\vec{P}/2$) times the volume of the
proton Fermi sea. Case~(1.2) is identical to the hole-hole phase space
at second order, and can be obtained from cases~(2.3) and~(2.4) (upon
exchanging $P/2$ and $k$, and integrating over $\cos\theta_{{\bf k},{\bf
P}}$ and $P$), which both give the result for case~(1.2) above.

The NN interaction matrix element in Eq.~\eqref{eq:NN_1st_start} is
expanded in partial waves, resulting in
\begin{align}
\frac{E_{\text{NN}}^{(1)}}{V} &= \frac{1}{8\pi^4} \int 
\limits_0^{\frac{k_F^n+k_F^p}{2}} 
dk \, k^2 \sum_{l,S,J, T, M_T} (2J+1) \nonumber \\[1mm]
&\quad\times f^{M_T}\hspace{-0.5mm}(k) \, \langle k| V_{l,l}^{J,S,M_T}|k \rangle \left(1-(-1)^{l+S+T}\right) \,,
\end{align}
where $f^{M_T=0} \equiv f^{np}$. Writing out the sum over isospin
states and neglecting the $pp$ contribution (see Eq.~\eqref{eq:smallx})
leads to the NN Hartree-Fock energy~\eqref{eq:NN_1st_contr}.

\subsection{Second-order NN contribution}
\label{app:2N_2nd}

We first expand the interaction matrix elements entering the
second-order NN contribution~\eqref{eq:NN_2nd_start} in partial
waves. This generalizes Ref.~\cite{Hebeler:2009iv} to arbitrary
isospin asymmetries. After expanding the angular parts in spherical
harmonics, taking $\vec{k}'$ along the $z$ axis, $\vec{k}$ in the
$x$-$z$ plane, inserting $(-1)^{l+S+T}$ for each antisymmetrizer,
and neglecting the $pp$ contributions, we have
\begin{align}
&\sum \limits_{S, M_S,M_S',T,M_T} \big| \bra{\vec{k} S M_S T M_T} \mathcal{A}_{12} V_\text{NN} \ket{\vec{k}' S M_S' T M_T} \big|^2 \nonumber \\
&= \sum_{L,S} \sum \limits_{J,l,l'} \sum \limits_{\widetilde{J},\widetilde{l},\widetilde{l}'} P_L(\cos \theta_{\vec{k},\vec{k'}}) (4\pi)^2 \, i^{(l-l'+\widetilde{l}-\widetilde{l}')} \, (-1)^{\widetilde{l}+l'+L} \nonumber \\
&\quad \times \clebschG{l}{\widetilde{l}'}{L}{0}{0}{0} \clebschG{l'}{\widetilde{l}}{L}{0}{0}{0}\sqrt{(2l+1)(2l'+1)(2\widetilde{l}+1)(2\widetilde{l}'+1)} \nonumber \\
&\quad \times (2J+1)(2\widetilde{J}+1)
\begin{Bmatrix}
l & S & J \\
\widetilde{J}&L&\widetilde{l}'
\end{Bmatrix}
\begin{Bmatrix}
J & S & l' \\
\widetilde{l}&L&\widetilde{J}
\end{Bmatrix} \nonumber \\
&\quad \times \bigg[ \bra{k}V_{l',l}^{J, S, M_T=-1}\ket{k'} \bra{k'}V_{\widetilde{l}',\widetilde{l}}^{\widetilde{J},S, M_T=-1}\ket{k} \nonumber \\
&\qquad \times \left(1-(-1)^{l+S+1}\right) \left(1-(-1)^{\widetilde{l}+S+1}\right) \nonumber \\
&\quad + \bra{k}V_{l',l}^{J, S, M_T=0}\ket{k'} \bra{k'}V_{\widetilde{l}',\widetilde{l}}^{\widetilde{J}, S, M_T=0}\ket{k} \nonumber \\
&\qquad \times \left(1-(-1)^{l+S}\right) \left(1-(-1)^{\widetilde{l}+S}\right) \nonumber \\
&\quad + \bra{k}V_{l',l}^{J, S, M_T=0}\ket{k'} \bra{k'}V_{\widetilde{l}',\widetilde{l}}^{\widetilde{J}, S, M_T=0}\ket{k} \nonumber \\
&\qquad \times \left(1-(-1)^{l+S+1}\right) \left(1-(-1)^{\widetilde{l}+S+1}\right) \bigg] \,.
\label{eq:2ndsum}
\end{align}

\begin{figure*}[t]
\centering
\subfloat[$\,$ \hspace{0.3cm} Case (2.1): \newline $0 \leqslant \frac{P}{2}\leqslant \frac{k_F^n-k_F^p}{2}$  and $k_F^p \leqslant \frac{P}{2} \,.$]{\includegraphics[width=0.32\textwidth,clip=]{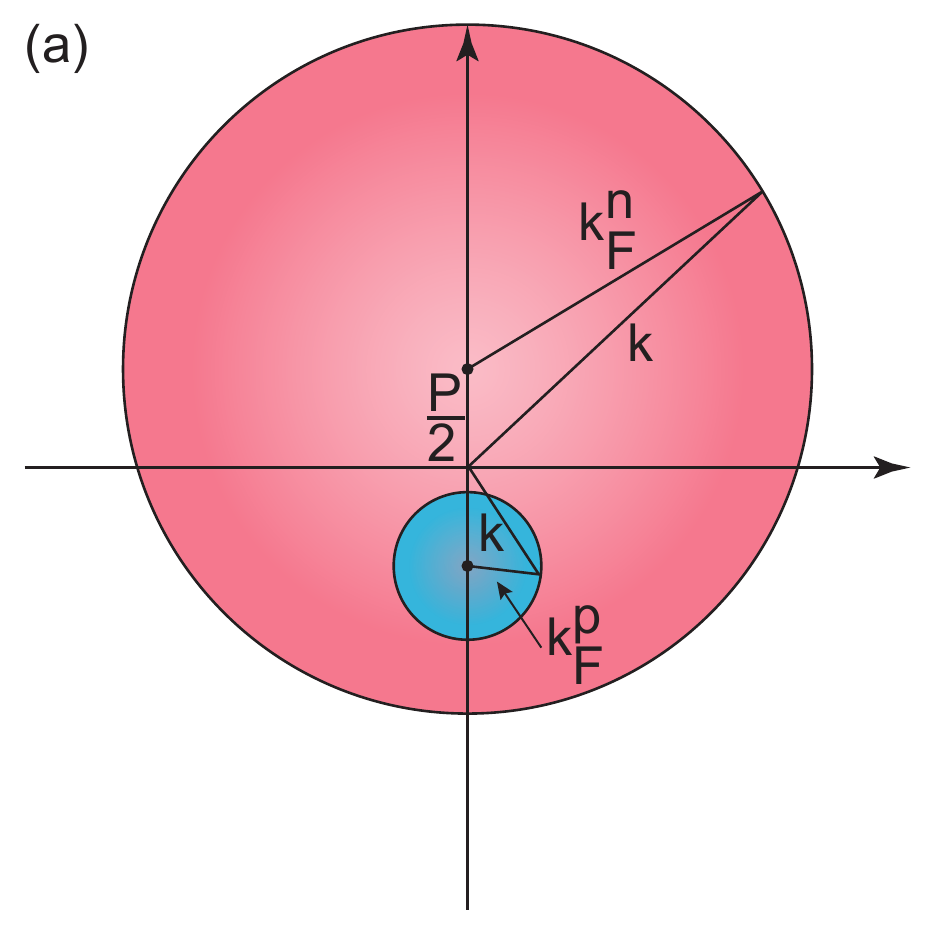}} \hfill
\subfloat[$\,$ \hspace{0.3cm} Case (2.2): \newline $0 \leqslant \frac{P}{2}\leqslant \frac{k_F^n-k_F^p}{2}$ and $k_F^p \geqslant \frac{P}{2} \,.$]{\includegraphics[width=0.32\textwidth,clip=]{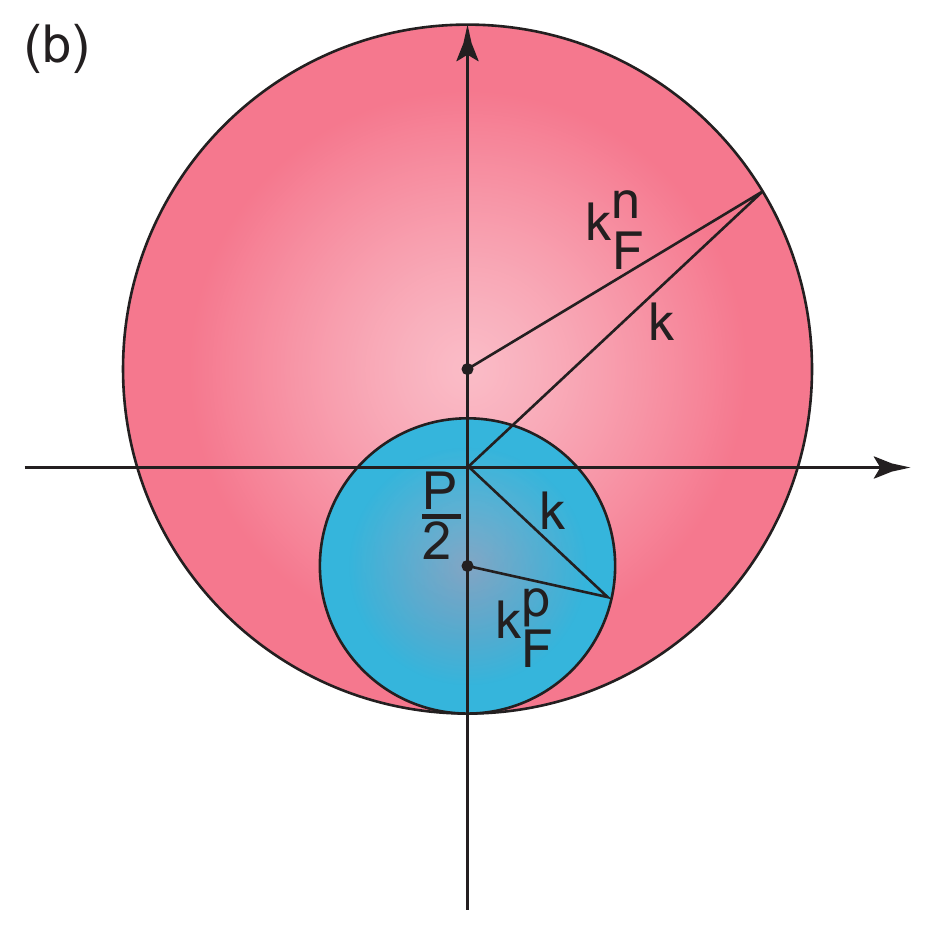}} \hfill
\subfloat[$\,$ \hspace{0.3cm} Case (2.3):\newline $\frac{k_F^n-k_F^p}{2}\leqslant \frac{P}{2}\leqslant \frac{k_F^n+k_F^p}{2}$ and $k_F^p\geqslant \frac{P}{2} \,.$]{\includegraphics[width=0.32\textwidth,clip=]{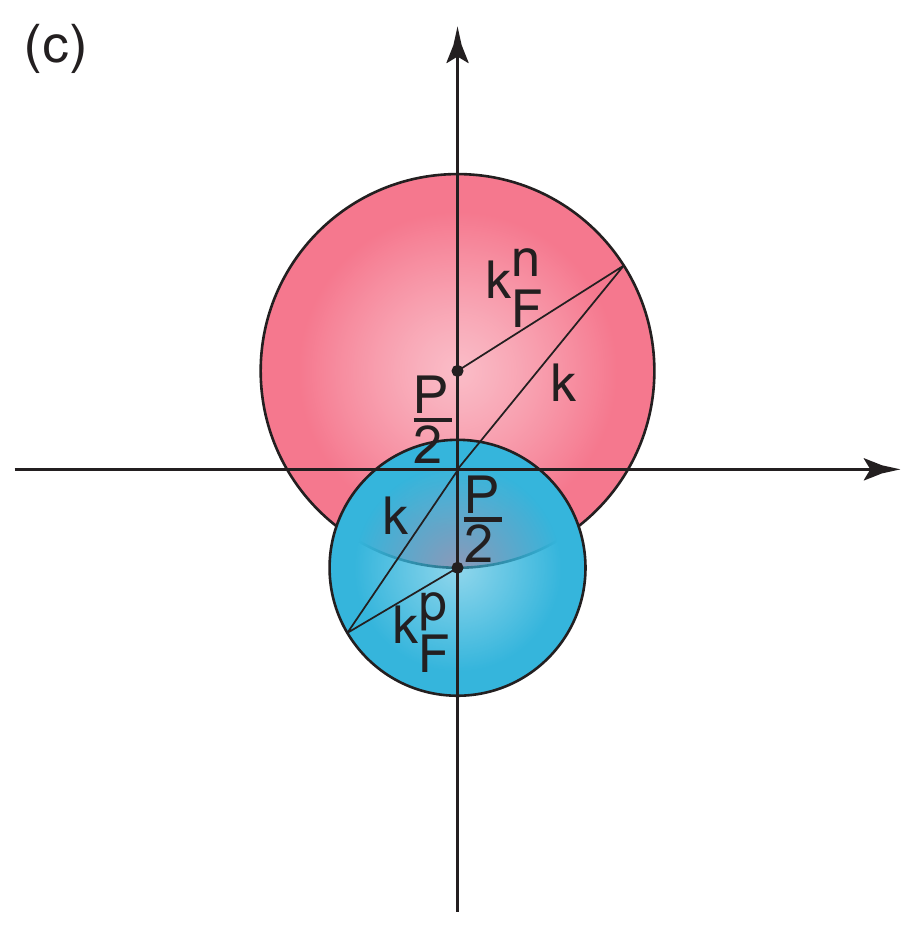}} \\
\subfloat[$\,$ \hspace{0.3cm} Case (2.4):\newline $\frac{k_F^n-k_F^p}{2}\leqslant \frac{P}{2}\leqslant \frac{k_F^n+k_F^p}{2}$ and $k_F^p\leqslant\frac{P}{2} \,.$]{\includegraphics[width=0.32\textwidth,clip=]{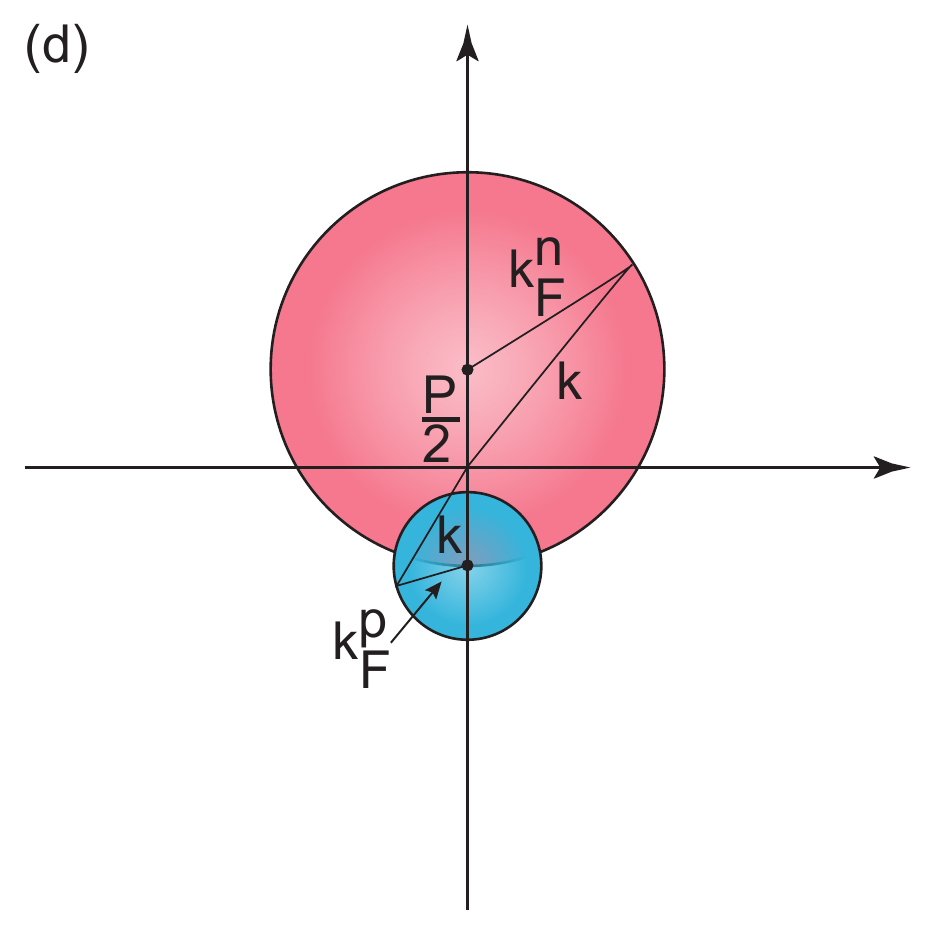}} \hfill
\caption{(Color online) Different regions contributing to the 
integral~\eqref{eq:Fhh}.  Red (blue) spheres represent the neutron
(proton) Fermi seas.\label{fig:2nd_spheres}}
\end{figure*}

\begin{figure*}[t]
\centering
\subfloat[$\,$ Case (3.1): $0 \leqslant \frac{P}{2}\leqslant \frac{k_F^n-k_F^p}{2}$]{\includegraphics[width=0.35\textwidth,clip=]{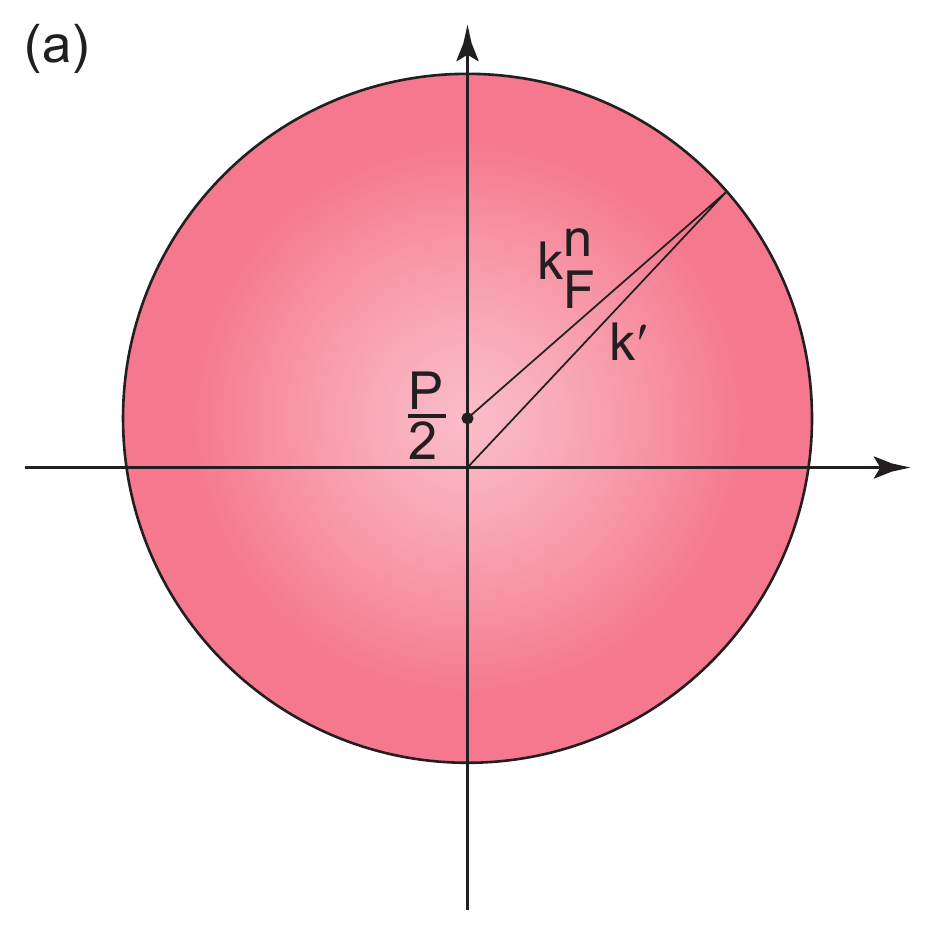}} \hfill
\subfloat[$\,$ Case (3.2): $\frac{k_F^n-k_F^p}{2} \leqslant \frac{P}{2}\leqslant \frac{k_F^n+k_F^p}{2}$]{\includegraphics[width=0.35\textwidth,clip=]{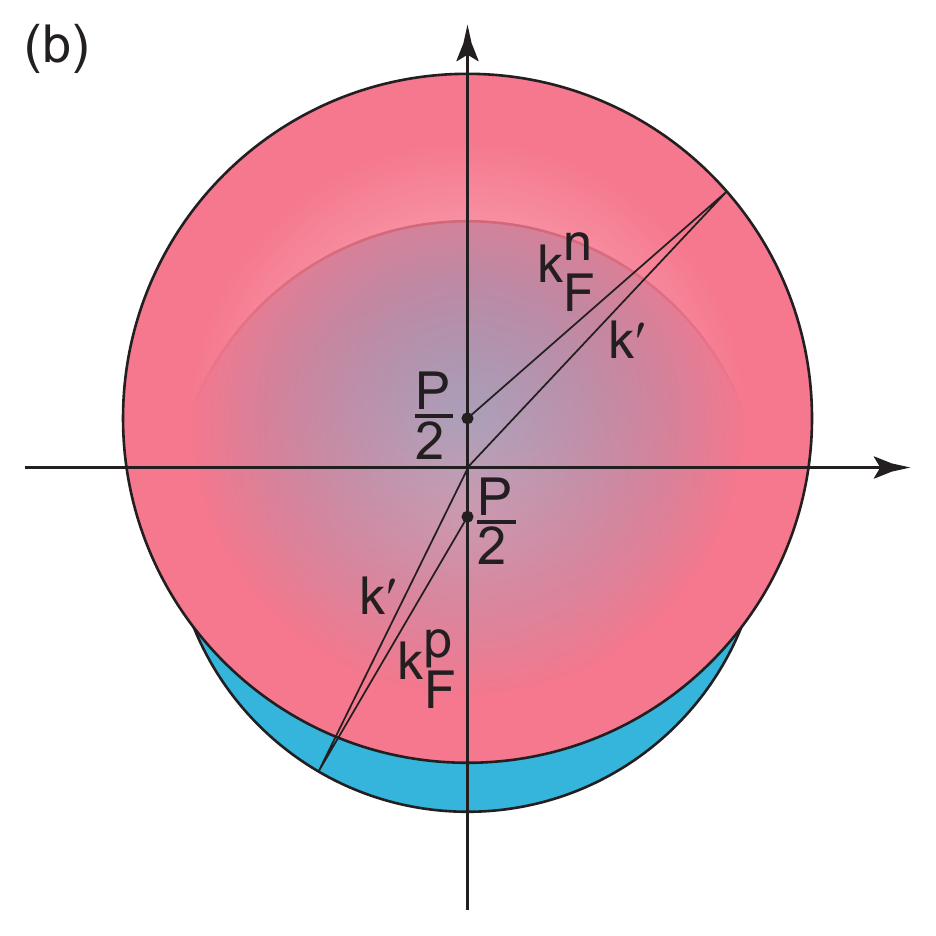}}
\caption{(Color online) Different regions contributing to the 
integral~\eqref{eq:Fpp}. Red (blue) spheres represent the neutron
(proton) Fermi surfaces.\label{fig:2nd_complements}}
\end{figure*}

Some of the integrals in Eq.~\eqref{eq:NN_2nd_start} can be performed
analytically. The angular integrations over the Fermi distribution
functions give rise to a function of the magnitude of the momenta,
\begin{align}
F^{np}(k,k',P) &= \int d\Omega_\vec{k} \int d\Omega_{\vec{k}'} 
\int d\Omega_\vec{P} \nonumber \\[1mm]
&\quad \times n^n_{\frac{\vec{P}}{2}+\vec{k}} \, n^p_{\frac{\vec{P}}{2}-\vec{k}} \, 
(1-n^n_{\frac{\vec{P}}{2}+\vec{k}'}) \, (1-n^p_{\frac{\vec{P}}{2}-\vec{k}'}) \,,
\end{align}
which is then used in Eq.~\eqref{eq:NN_2nd_contr}. Again, we focus
on the $np$ case. To derive $F^{np}(k,k',P)$, let us take $\vec{P}$
along the $z$ axis and $\vec{k}$ in the $x$-$z$ plane. We consider
only the $L=0$ contribution in the partial-wave
expression~\eqref{eq:2ndsum}, which is equivalent to angle
averaging. In this approximation, the $\varphi_{\vec{k}'}$ integration
yields $2\pi$ and we are left with
\begin{align}
F^{np}(k,k',P) &= 16 \, \pi^3  \int \limits_{-1}^{1} d\cos  \theta_{\vec{k},\vec{P}} \int \limits_{-1}^{1} d\cos \theta_{\vec{k}',\vec{P}} \nonumber \\
&\quad \times n^n_{\frac{\vec{P}}{2}+\vec{k}} \, n^p_{\frac{\vec{P}}{2}-\vec{k}} \, 
(1-n^n_{\frac{\vec{P}}{2}+\vec{k}'}) \, (1-n^p_{\frac{\vec{P}}{2}-\vec{k}'}) \,.
\label{eq:FFF}
\end{align}
The two integrals can be worked out separately, giving rise to two
functions that account for the hole-hole (hh) and particle-particle
(pp) phase space
\begin{equation}
F^{np}(k, k', P) = 16 \, \pi^3 F^{np}_{\text{hh}}(k, P) \, 
F^{np}_{\text{pp}}(k', P) \,.
\end{equation}
Let us start with the hole-hole part. This is given by the volume of
the intersection of two Fermi spheres with radii $k_F^n$ and $k_F^p$
whose centers are displaced by $\vec{P}$. Depending on the value of $P$,
one has to distinguish four different cases, which are shown in
Fig.~\ref{fig:2nd_spheres},
\begin{itemize}
\item[(2.1)] $\displaystyle 0 \leqslant \frac{P}{2} \leqslant \frac{k_F^n-k_F^p}{2} \quad$ and $\displaystyle \quad k_F^p \leqslant \frac{P}{2} \,;$
\item[(2.2)] $\displaystyle 0 \leqslant \frac{P}{2} \leqslant \frac{k_F^n-k_F^p}{2} \quad$ and $\displaystyle \quad k_F^p \geqslant \frac{P}{2} \,;$
\item[(2.3)] $\displaystyle \frac{k_F^n-k_F^p}{2} \leqslant \frac{P}{2} \leqslant \frac{k_F^n+k_F^p}{2} \quad$ and $\displaystyle \quad k_F^p \geqslant \frac{P}{2} \,;$
\item[(2.4)] $\displaystyle \frac{k_F^n-k_F^p}{2} \leqslant \frac{P}{2} \leqslant \frac{k_F^n+k_F^p}{2} \quad$ and $\displaystyle \quad k_F^p \leqslant \frac{P}{2} \,.$
\end{itemize}
It is useful to express the function $F^{np}_{\text{hh}}(k,P)$ as
\begin{equation}
F^{np}_{\text{hh}}(k,P) = \int \limits_{f_1(k,P)}^{f_2(k,P)} d\cos \theta_{\vec{k},\vec{P}} \ n^n_{\frac{\vec{P}}{2}+\vec{k}} \, n^p_{\frac{\vec{P}}{2}-\vec{k}} \,,
\label{eq:Fhh}
\end{equation}
where the lower and upper limits of the integration will be different
in each case. In the first two total-overlap cases, one has $f_1(k,P)
= -1$, and for case (2.1)
\begin{equation}
f_2(k,P) = \begin{cases} 
-1 & k \leqslant \frac{P}{2}-k_F^p \,, \\
\frac{\left( k_F^p\right)^2-\left(\frac{P}{2}\right)^2-k^2}{2k\frac{P}{2}} & \frac{P}{2}-k_F^p \leqslant k \leqslant k_F^p+\frac{P}{2} \,, \\
-1 & k \geqslant k_F^p+\frac{P}{2} \,;
\end{cases}
\end{equation}
while for case (2.2)
\begin{equation}
f_2(k,P) = \begin{cases} 
1 & k\leqslant k_F^p-\frac{P}{2} \,, \\
\frac{\left( k_F^p\right)^2-\left(\frac{P}{2}\right)^2-k^2}{2k\frac{P}{2}} & k_F^p- \frac{P}{2} \leqslant k\leqslant k_F^p+\frac{P}{2} \,, \\
-1 & k \geqslant k_F^p+\frac{P}{2} \,.
\end{cases}
\end{equation}
The partial overlap cases yield more involved integration limits. We
find for case (2.3):
\begin{equation}
f_1(k,P) = \begin{cases}
-1 & k\leqslant k_F^n- \frac{P}{2} \,, \\
\frac{\left( k_F^n\right)^2-\left(\frac{P}{2}\right)^2-k^2}{-2k\frac{P}{2}} & k_F^n- \frac{P}{2}\leqslant k \leqslant k_0 \,, \\
-1 & k \geqslant k_0 \,;
\end{cases}
\end{equation}
and
\begin{equation}
f_2(k,P) = \begin{cases} 
1 & k\leqslant k_F^p- \frac{P}{2} \,, \\
\frac{\left( k_F^p\right)^2-\left(\frac{P}{2}\right)^2-k^2}{2k\frac{P}{2}} &  k_F^p- \frac{P}{2}\leqslant k \leqslant k_0 \,, \\
-1 & k \geqslant k_0 \,;
\end{cases}
\end{equation}
where $k_0=\sqrt{\frac{\left( k_F^n \right)^2+\left(k_F^p
\right)^2}{2}-\left( \frac{P}{2} \right)^2}$. Case (2.4) gives
\begin{equation}
f_1(k,P) = \begin{cases} 
-1 & k \leqslant k_F^n-\frac{P}{2} \,, \\
\frac{\left( k_F^n\right)^2-\left(\frac{P}{2}\right)^2-k^2}{-2k\frac{P}{2}} & k_F^n-\frac{P}{2} \leqslant k \leqslant k_0 \,, \\
-1 & k \geqslant k_0 \,;
\end{cases}
\end{equation}
and
\begin{equation}
f_2(k,P) = \begin{cases} 
-1 & k\leqslant \frac{P}{2}-k_F^p  \,, \\
\frac{\left( k_F^p\right)^2-\left(\frac{P}{2}\right)^2-k^2}{2k\frac{P}{2}} & \frac{P}{2}-k_F^p \leqslant k \leqslant k_0\,, \\
-1 & k \geqslant k_0 \,.
\end{cases}
\end{equation}

The second integral in Eq.~\eqref{eq:FFF} is performed similarly, with
the difference that now the volume excluded by the union of the two
Fermi spheres contributes. One can distinguish two cases
\begin{itemize}
\item[(3.1)] $\displaystyle 0 \leqslant \frac{P}{2} \leqslant \frac{k_F^n-k_F^p}{2} \,,$
\item[(3.2)] $\displaystyle \frac{k_F^n-k_F^p}{2} \leqslant \frac{P}{2} \leqslant \frac{k_F^n+k_F^p}{2} \,,$
\end{itemize}
which are shown in Fig.~\ref{fig:2nd_complements}. As for the
hole-hole cases, we express the function $F^{np}_{\text{pp}}(k',P)$ as
\begin{equation}
F^{np}_{\text{pp}}(k',P) = \int \limits_{f_1(k',P)}^{f_2(k',P)} d\cos \theta_{\vec{k}',\vec{P}} \ (1-n^n_{\frac{\vec{P}}{2}+\vec{k}'}) \, (1-n^p_{\frac{\vec{P}}{2}-\vec{k}'}) \,.
\label{eq:Fpp}
\end{equation}
In the total overlap case (3.1), we have $f_1(k',P) = -1$ and
\begin{equation}
f_2(k',P) = \begin{cases} 
-1 & k' \leqslant k_F^n - \frac{P}{2} \,, \\
\frac{\left( k_F^n\right)^2-\left(\frac{P}{2}\right)^2-k'^2}{-2k'\frac{P}{2}} & 
k_F^n - \frac{P}{2} \leqslant k' \leqslant k_F^n + \frac{P}{2} \,, \\
1 & k'\geqslant k_F^n + \frac{P}{2} \,. \\
\end{cases}
\end{equation}
The partial overlap case (3.2) yields
\begin{equation}
f_1(k',P) = \begin{cases}
-1 & k'\leqslant k_0 \,, \\
\frac{\left( k_F^p\right)^2-\left(\frac{P}{2}\right)^2-k'^2}{2k'\frac{P}{2}} & k_0\leqslant k'\leqslant k_F^p +\frac{P}{2} \,, \\
-1 & k'\geqslant k_F^p+\frac{P}{2} \,;
\end{cases}
\end{equation}
and
\begin{equation}
f_2(k',P) = \begin{cases} 
-1 & k'\leqslant k_0 \,, \\
\frac{\left( k_F^n\right)^2-\left(\frac{P}{2}\right)^2-k'^2}{-2k'\frac{P}{2}} & k_0 \leqslant k' \leqslant\frac{P}{2}+k_F^n \,, \\
1 & k'\geqslant k_F^n+\frac{P}{2} \,.
\end{cases}
\end{equation}

\section{First-order 3N contribution}
\label{app:3N_1st}

Next, we discuss the contributions from N$^2$LO 3N forces
$V_\text{3N}=V_c+V_D+V_E$ and calculate the Hartree-Fock energy
density~\eqref{eq:3N_1st_start}. The different 3N interaction parts
read~\cite{vanKolck:1994yi,Epelbaum:2002vt}
\begin{align}
V_c &= \frac{1}{2} \biggl( \frac{g_A}{2F_\pi} \biggr)^2 \sum \limits_{i \neq j \neq k} \frac{(\bm{\sigma}_i \cdot \vec{q}_i)(\bm{\sigma}_j \cdot \vec{q}_j)}{(q_i^2+m_\pi^2)(q_j^2+m_\pi^2)} \, F_{ijk}^{\alpha \beta} \, \tau_i^\alpha \, \tau_j^\beta \,, \label{eq:VC_definition} \\
V_D &= -\frac{g_A}{8F_\pi^2} \frac{c_D}{F_\pi^2 \Lambda_\chi} \sum \limits_{i \neq j \neq k} \frac{\bm{\sigma}_j \cdot \vec{q}_j}{q_j^2+m_\pi^2} (\bm{\sigma}_i \cdot \vec{q}_j)(\bm{\tau}_i \cdot \bm{\tau}_j) \,,
\label{eq:VD_definition} \\
V_E &= \frac{c_E}{2F_\pi^4 \Lambda_\chi} \sum \limits_{j \neq k} \, (\bm{\tau}_j \cdot \bm{\tau}_k ) \,,
\label{eq:VE_definition}
\end{align}
with $g_A=1.29$, $F_\pi=92.4 \, {\rm MeV}$, $m_\pi=138.04 \, {\rm
MeV}$, and $\Lambda_\chi=700 \, {\rm MeV}$. $\vec{q}_i = \vec{k}_i'
-\vec{k}_i$ is the difference of initial and final nucleon momenta and
\begin{align}
F_{ijk}^{\alpha \beta} &= \delta^{\alpha \beta} \left[ -\frac{4c_1m_\pi^2}{F_\pi^2}+\frac{2c_3}{F_\pi^2} \, \vec{q}_i \cdot \vec{q}_j \right] \nonumber \\[1mm]
&\quad + \sum_\gamma \frac{c_4}{F_\pi^2} \, \epsilon^{\alpha \beta \gamma} \, \tau_k^\gamma \, \bm{\sigma}_k \cdot (\vec{q}_i \times \vec{q}_j) \,.
\end{align}
We consider the different 3N contributions for the $nnp$ case
according to the approximation~\eqref{eq:smallx}. The $nnn$
expressions are given in Ref.~\cite{Hebeler:2009iv}.

\subsection{$V_c$ contribution}

Let us write Eq.~\eqref{eq:VC_definition} as
\begin{equation}
V_c = \frac{1}{2} \biggl( \frac{g_A}{2 F_\pi} \biggr)^2 \biggl( G^{(1)} + \frac{c_4}{F_\pi^2} \, G^{(2)} \biggr) \,,
\end{equation}
with
\begin{align}
G^{(1)} &= \sum_{i \neq j \neq k} f_{ij} \, (\bm{\tau}_i \cdot \bm{\tau}_j) \,, \\
G^{(2)} &= \sum_{i \neq j \neq k} g_{ij} \, \bm{\tau}_k \cdot (\bm{\tau}_i \times \bm{\tau}_j) \, \bm{\sigma}_k\cdot(\vec{q}_i \times \vec{q}_j) \,,
\end{align}
and
\begin{align}
f_{ij} &= \frac{(\bm{\sigma}_i \cdot \vec{q}_i)(\bm{\sigma}_j \cdot \vec{q}_j)}{(q_i^2+m_\pi^2)(q_j^2+m_\pi^2)} \left[ -\frac{4c_1m_\pi^2}{F_\pi^2}+\frac{2c_3}{F_\pi^2}\,  \vec{q}_i \cdot \vec{q}_j \right] \,, \\
g_{ij} &= \frac{(\bm{\sigma}_i \cdot \vec{q}_i)(\bm{\sigma}_j \cdot \vec{q}_j)}{(q_i^2+m_\pi^2)(q_j^2+m_\pi^2)} \,.
\end{align}
We need to calculate the matrix element $\langle 123| \mathcal{A}_{123}
V_{c}|123\rangle$, with three-body antisymmetrizer
\begin{equation} 
\mathcal{A}_{123} = 1-P_{12}-P_{13}-P_{23}+P_{12}P_{23}+P_{13}P_{23} \,,
\end{equation}
where the particle-exchange operator acts on momentum, spin, and isospin
$P_{ij} = P_{ij}^{k} \, P_{ij}^{\sigma} \, P_{ij}^{\tau}$. We first consider
the isospin-exchange operators $P_{ij}^\tau$ for the $G^{(1)}$ part
\begin{align}
\langle nnp | \mathcal{A}_{123} G^{(1)}& |nnp \rangle = \langle nnp | \sum_{i \neq j \neq k} f_{ij} (\bm{\tau}_i \cdot \bm{\tau}_j) |nnp \rangle \nonumber \\ 
&\quad- \langle nnp | P_{12}^{\sigma k} \sum_{i \neq j \neq k} f_{ij} (\bm{\tau}_i \cdot \bm{\tau}_j) |nnp \rangle \nonumber \\
&\quad- \langle pnn | P_{13}^{\sigma k} \sum_{i \neq j \neq k} f_{ij} (\bm{\tau}_i \cdot \bm{\tau}_j) |nnp \rangle \nonumber \\
&\quad- \langle npn | P_{23}^{\sigma k} \sum_{i \neq j \neq k} f_{ij} (\bm{\tau}_i \cdot \bm{\tau}_j) |nnp \rangle \nonumber \\
&\quad+ \langle npn | P_{12}^{\sigma k}P_{23}^{\sigma k} \sum_{i \neq j \neq k} f_{ij} (\bm{\tau}_i \cdot \bm{\tau}_j)|nnp \rangle \nonumber \\
&\quad+ \langle pnn | P_{13}^{\sigma k}P_{23}^{\sigma k} \sum_{i \neq j \neq k} f_{ij} (\bm{\tau}_i \cdot \bm{\tau}_j) |nnp \rangle \,.
\end{align}
Evaluating the matrix elements for the different $\bm{\tau}_i \cdot
\bm{\tau}_j$, we find
\begin{align}
&\langle nnp | \mathcal{A}_{123} G^{(1)} |nnp \rangle \nonumber \\[1mm]
&=2 \Big[ (f_{12}-f_{13}-f_{23}) 
- P_{12}^{\sigma k} (f_{12}-f_{13}- f_{23}) \nonumber \\[1mm]
&\quad - 2P_{13}^{\sigma k}f_{13} -2 P_{23}^{\sigma k}f_{23} +2P_{12}^{\sigma k}P_{23}^{\sigma k}f_{23} + 2P_{13}^{\sigma k}P_{23}^{\sigma k}f_{13} \Big] \,.
\label{eq:VC_partial_result_G1}
\end{align}
In the same way the $G^{(2)}$ part yields matrix elements of triple
products, $\langle nnp | \bm{\tau}_1 \cdot (\bm{\tau}_2 \times
\bm{\tau}_3) | nnp \rangle$, and permutations thereof. These can
be evaluated using, for example,
\begin{align}
\langle nnp | \epsilon^{\alpha \beta \gamma}\tau_1^\alpha\tau_2^\beta\tau_3^\gamma |nnp \rangle &= \langle nnp | \epsilon^{zz\gamma}\tau_1^z\tau_2^z\tau_3^\gamma |nnp \rangle = 0 \,, \\
\langle npn | \epsilon^{\alpha \beta \gamma}\tau_1^\alpha\tau_2^\beta\tau_3^\gamma |nnp \rangle &= \langle npn | \epsilon^{z\beta \gamma}\tau_1^z\tau_2^\beta\tau_3^\gamma |nnp \rangle = -2i \,.
\end{align}

We then consider the spin-exchange part. The spin-exchange operator is
given by $P_{ij}^\sigma=(1+\bm{\sigma}_i \cdot \bm{\sigma}_j)/2$. When
summing over spins, only terms without Pauli matrices give
non-vanishing contributions. For example, for the $f_{ij}$ part, this
leaves terms like
\begin{align}
(\sigma_1^a \sigma_2^a ) (\sigma_1^b q_1^b ) (\sigma_2^c q_2^c ) 
&= (\delta^{ab}+i\epsilon^{abd} \sigma_1^d)(\delta^{ac}+i\epsilon^{ace}\sigma_2^e) q_1^b q_2^c \nonumber \\ 
&\overset{\text{Tr}}{\longrightarrow} 8 \, \delta^{bc}q_1^b q_2^c = 8 \, \vec{q}_1 \cdot \vec{q}_2 \,,
\end{align}
where the second line is given after tracing over the three spins in
Eq.~\eqref{eq:3N_1st_start}. For the same reason, this leaves for the
$g_{ij}$ part terms like
\begin{align}
&(\sigma_1^a\sigma_2^a)(\sigma_2^b\sigma_3^b)(\sigma_1^c q_1^c)(\sigma_2^d q_2^d) \, \sigma_3^e (\vec{q}_1 \times \vec{q}_2)^e \nonumber \\
\overset{\text{Tr}}{\longrightarrow} &- 8 \, i \epsilon^{cdb} \, q_1^c \, q_2^d \, (\vec{q}_1 \times \vec{q}_2)^b = - 8 \, i (\vec{q}_1 \times \vec{q}_2)^2 \,.
\end{align}
We then apply the momentum-exchange operator and evaluate
$\vec{q}_i=\vec{k}_i'-\vec{k}_i$, where $\vec{k}_i'$ corresponds to
the bra and $\vec{k}_i$ to the ket state. As a result, the $V_c$
contribution to the Hartree-Fock energy density~\eqref{eq:3N_1st_start}
is given by
\begin{widetext}
\begin{align}
\frac{E_{V_c}^{(1)}}{V} \Bigg|_{nnp} &= \frac{4}{3} \biggl( \frac{g_A}{2 F_\pi} \biggr)^2 \int \frac{d \vec{k}_1 d \vec{k}_2 d \vec{k}_3}{(2 \pi)^9} \, n_{\vec{k}_1}^n n_{\vec{k}_2}^n n_{\vec{k}_3}^p \, f_\text{R}^2 \nonumber \\ 
&\quad \times \Bigg( -\frac{4c_1 m_\pi^2}{F_\pi^2} \left[ \frac{k_{12}^2}{2(k_{12}^2+m_{\pi}^2)^2}+\frac{k_{23}^2}{(k_{23}^2+m_\pi^2)^2}+ \frac{k_{13}^2}{(k_{13}^2+m_\pi^2)^2}-\frac{\vec{k}_{23} \cdot \vec{k}_{31}}{(k_{23}^2+m_\pi^2)(k_{13}^2+m_\pi^2)} \right] \nonumber \\[1mm]
&\quad- \frac{2c_3}{F_\pi^2} \left[ \frac{k_{12}^4}{2(k_{12}^2+m_{\pi}^2)^2}+\frac{k_{23}^4}{(k_{23}^2+m_\pi^2)^2}+ \frac{k_{13}^4}{(k_{13}^2+m_\pi^2)^2}-\frac{(\vec{k}_{23} \cdot \vec{k}_{31})^2}{(k_{23}^2+m_\pi^2)(k_{13}^2+m_\pi^2)} \right] \nonumber \\[1mm]
&\quad- \frac{c_4}{F_\pi^2} \left[ \frac{(\vec{k}_{12} \times \vec{k}_{23})^2}{(k_{12}^2+m_\pi^2)(k_{23}^2+m_\pi^2)}+\frac{(\vec{k}_{12} \times \vec{k}_{31})^2}{(k_{12}^2+m_\pi^2)(k_{31}^2+m_\pi^2)} +\frac{(\vec{k}_{23} \times \vec{k}_{31})^2}{(k_{23}^2+m_\pi^2)(k_{31}^2+m_\pi^2)} \right] \Bigg) \,.
\label{eq:3N_VC_contr}
\end{align}
\end{widetext}

\subsection{$V_D$ contribution}

To calculate $\langle 123| \mathcal{A}_{123} V_{D}|123\rangle$, we
first consider the isospin part, which is of the $G^{(1)}$ form. Using
the results from Eq.~\eqref{eq:VC_partial_result_G1}, we find for the
matrix element, dropping terms that give non-vanishing contributions
after summing over spins,
\begin{align}
\bra{nnp} \mathcal{A}_{123} V_{D} \ket{nnp} 
&= 2 \Big[ -P_{12}^{\sigma k} d_{12}-2P_{13}^{\sigma k} d_{13} -2 P_{23}^{\sigma k} d_{23} \nonumber \\[1mm]
&\quad+ 2 P_{13}^{\sigma k}P_{23}^{\sigma k} d_{23} + 2 P_{13}^{\sigma k}P_{23}^{\sigma k} d_{13} \Big] \,,
\end{align}
where
\begin{equation}
d_{ij} = -\frac{g_A}{8F_\pi^2} \frac{c_D}{F_\pi^2 \Lambda_\chi} \frac{\bm{\sigma}_j \cdot \vec{q}_j}{q_j^2+m_\pi^2} (\bm{\sigma}_i \cdot \vec{q}_j ) \,.
\end{equation}
Summing over spins leaves terms like
\begin{equation}
\frac{1}{4} (1+\bm{\sigma}_1 \cdot \bm{\sigma}_2)(1+\bm{\sigma}_2 \cdot \bm{\sigma}_3) d_{23} \overset{\text{Tr}}{\longrightarrow} 2 \, \vec{q}_2^2 \,.
\end{equation}
Finally, evaluating the momentum-exchange operators, we find for the
$V_D$ contribution to the Hartree-Fock energy density
\begin{align}
\frac{E_{V_D}^{(1)}}{V}\Bigg|_{nnp} &= \frac{g_A}{6 F_\pi^2} \frac{c_D}{F_\pi^2 \Lambda_{\chi}} \int \frac{d \vec{k}_1 d \vec{k}_2 d \vec{k}_3}{(2 \pi)^9} \, n_{\vec{k}_1}^n n_{\vec{k}_2}^n n_{\vec{k}_3}^p \, f_\text{R}^2 \nonumber \\ 
&\quad\times \left[\frac{k_{12}^2}{k_{12}^2+m_{\pi}^2} + \frac{k_{23}^2}{k_{23}^2+m_{\pi}^2} +\frac{k_{13}^2}{k_{13}^2+m_{\pi}^2} \right] \,.
\label{eq:3N_VD_contr}
\end{align}

\subsection{$V_E$ contribution}

The isopin part of the matrix element $\langle 123| \mathcal{A}_{123}
V_{E}|123\rangle$ is also of the $G^{(1)}$ form. Using the results
from Eq.~\eqref{eq:VC_partial_result_G1} with $f_{ij}=1$, we have
\begin{align}
\langle nnp| \mathcal{A}_{123} \sum \limits_{j \neq k} \bm{\tau}_j \cdot \bm{\tau}_k |nnp \rangle &= 2 \Bigl[ -1 +P_{12}^{\sigma k}-2P_{13}^{\sigma k}-2P_{23}^{\sigma k} \nonumber \\
&\quad +2P_{12}^{\sigma k}P_{23}^{\sigma k}+2P_{13}^{\sigma k}P_{23}^{\sigma k} \Bigr] \,.
\end{align}
Summing over spins, only the $1/2$ part of the spin-exchange operator
$P_{ij}^{\sigma}=(1+\bm{\sigma}_i \cdot \bm{\sigma}_j)/2$ gives
non-vanishing contributions, so that the matrix element yields $-24$
after the spin traces. As a result, the $V_E$ contribution to the
Hartree-Fock energy density is given by
\begin{equation}
\frac{E_{V_E}^{(1)}}{V} \Bigg|_{nnp} = -2 \frac{c_E}{F_\pi^4 \Lambda_{\chi}}
\int \frac{d \vec{k}_1 d \vec{k}_2 d \vec{k}_3}{(2 \pi)^9} \, 
n_{\vec{k}_1}^n n_{\vec{k}_2}^n n_{\vec{k}_3}^p \, f_\text{R}^2 \,.
\label{eq:3N_VE_contr}
\end{equation}

\bibliography{literature}
\end{document}